\documentclass[12pt]{article}
\RequirePackage[OT1]{fontenc}
\RequirePackage{amsthm,amsmath,amsfonts}
\usepackage[authoryear]{natbib}
\RequirePackage[colorlinks,citecolor=blue,urlcolor=blue]{hyperref}
\usepackage[a4paper, total={6.5in, 8.5in}]{geometry}
\usepackage{latexsym,graphicx,subcaption,verbatim,nameref}
\usepackage{diagbox}
\usepackage{amsfonts,amsmath,amscd,amssymb}
\usepackage{latexsym,graphicx,subcaption,verbatim,nameref}
\usepackage[affil-it]{authblk}
\usepackage[all,cmtip]{xy}
\usepackage[toc,title,titletoc]{appendix}
\usepackage[capposition=top]{floatrow}
\usepackage{rotating}
\usepackage{threeparttable,booktabs}
\usepackage{makecell}
\usepackage{csquotes}
\usepackage{chngcntr}
\usepackage{graphicx}
\usepackage{esint}
\usepackage{relsize}
\usepackage{color}
\usepackage{multirow}
\usepackage{algcompatible}
\usepackage{algorithm}
\usepackage{algpseudocode}
\usepackage{setspace}
\usepackage{mathtools}

\newcommand{\expect}{\operatorname{E}\expectarg}
\DeclarePairedDelimiterX{\expectarg}[1]{(}{)}{%
  \ifnum\currentgrouptype=16 \else\begingroup\fi
  \activatebar#1
  \ifnum\currentgrouptype=16 \else\endgroup\fi
}

\newcommand{\prob}{\operatorname{Pr}\expectarg}
\DeclarePairedDelimiterX{\probarg}[1]{(}{)}{%
  \ifnum\currentgrouptype=16 \else\begingroup\fi
  \activatebar#1
  \ifnum\currentgrouptype=16 \else\endgroup\fi
}

\newcommand{\innermid}{\nonscript\;\delimsize\vert\nonscript\;}
\newcommand{\activatebar}{%
  \begingroup\lccode`\~=`\|
  \lowercase{\endgroup\let~}\innermid 
  \mathcode`|=\string"8000
}
\onehalfspace
\numberwithin{figure}{section}
\numberwithin{table}{section}
\numberwithin{equation}{section}
\theoremstyle{plain}
\newtheorem{theorem}{Theorem}[section]

\newtheorem{lemma}[theorem]{Lemma}
\theoremstyle{definition}
\newtheorem{definition}[theorem]{Definition}
\newtheorem{remark}[theorem]{Remark}
\newtheorem{example}[theorem]{Example}

\newcommand*{\Scale}[2][4]{\scalebox{#1}{$#2$}}%
\newcommand{\varA}[1]{{\operatorname{#1}}}
\newcommand{\varB}[1]{{\operatorname{\mathit{#1}}}}
\newcommand{\algrule}[1][.2pt]{\par\vskip.5\baselineskip\hrule height #1\par\vskip.5\baselineskip}

\providecommand{\keywords}[1]{\textbf{\textit{Keywords---}} #1}
\let\oldref\ref
\renewcommand{\ref}[1]{(\oldref{#1})}

\begin{document}

\title{A simulation-based approach to estimate joint model of longitudinal and event-time data with many missing longitudinal observations 
}
\author[1]{Yanqiao Zheng%
\thanks{Email: \texttt{yanqiaoz@buffalo.edu}}}

\author[2]{Xiaobing Zhao%
\thanks{Email: \texttt{maxbzhao@126.com}; Corresponding Author. Address: No. 18, Xueyuan Street, Xiasha Higher Education Park, Hangzhou, Zhejiang, 310018, China}}

\author[1]{Xiaoqi Zhang%
\thanks{Email: \texttt{xiaoqizh@buffalo.edu}}}
\affil[1]{School of Finance\\ Zhejiang University of Finance and Economics}
\affil[2]{School of Data Science\\ Zhejiang University of Finance and Economics}
\maketitle

\newpage
\begin{abstract}
Joint models of longitudinal and event-time data have been extensively studied and applied in many different fields. Estimation of joint models is challenging, most present procedures are computational expensive and have a strict requirement on data quality. In this study, a novel simulation-based procedure is proposed to estimate a general family of joint models, which include many widely-applied joint models as special cases. Our procedure can easily handle low-quality data where longitudinal observations are systematically missed for some of the covariate dimensions. In addition, our estimation procedure is compatible with parallel computing framework when combining with stochastic descending algorithm, it is perfectly applicable to massive data and therefore suitable for many financial applications. Consistency and asymptotic normality of our estimator are proved,  a simulation study is conducted to illustrate its effectiveness. Finally, as an application, the procedure is applied to estimate pre-payment probability of a massive consumer-loan dataset drawn from one biggest P2P loan platform of China. \end{abstract}

\section{Introduction}\label{introduction}

Joint models of longitudinal and event-time data have been extensively studied. It has the following general form:

\begin{equation}\label{joint model}
\begin{aligned}
&\lambda\left(t,Z(t)\right)=\lambda_{0}
\left(t\right)
\exp\left(b^{\top}Z(t)\right)\\
&Z\left(t\right)=\mathcal{Z}\left(t,a\right)
\end{aligned}
\end{equation}
where $\lambda\left(t,z\right)$ is the conditional hazard function given time $t$ and the temporal covariate $Z(t)=z$, which has the Cox form and consists of a non-parametric baseline hazard $\lambda_0(t)$ and a parametric component, $\exp\left(b^{\top}z\right)$. The temporal covariate is given through a $\mathfrak{p}$-dimensional stochastic process $\mathcal{Z}$ which is parametrized by a set of parameter $a$.

Joint models of type \eqref{joint model} have many variants, which are derived either from changing the functional form of hazard function or from different specification of longitudinal processes. A variety of alternative functional forms of hazard have been proposed and widely studied in literature \citep{rizopoulos2011dynamic,kim2013joint,
wang2001jointly,tsiatis1995modeling,
taylor1994stochastic,chen2014joint}, although they are quite useful in different application settings, from the perspective of estimation strategy, they do not really make a difference to the ordinary form of Cox hazard function \eqref{joint model}.
In contrast, variations in specification of longitudinal process is more interesting, as different specifications associate with different protocol models in survival analysis, to which different estimation procedures have to be developed. 

For instance, when 
\begin{equation}\label{stationary cox}
\mathcal{Z}\left(t,a\right)\equiv z
\end{equation}
the longitudinal process reduces to a deterministic constant process ($z$), model \eqref{joint model} reduces to the standard Cox proportional hazard model \citep{cox1972regression} which can be estimated through the classical maximum partial likelihood procedure(MPL) \citep{cox1972regression,andersen1982cox,andersen1992repeated}.

When longitudinal process is given through a linear mixed model as below:
\begin{equation}\label{bio-stat longi}
\mathcal{Z}\left(t,a\right)=\alpha\cdot \mathcal{Z}_1(t)+\beta\cdot \mathcal{Z}_2(t)+e
\end{equation}
model \eqref{joint model} becomes a family of joint models that are extensively applied by bio-statisticians in medical-related fields where clinical-trail data is available \citep{taylor1994stochastic,tsiatis1995modeling,wang2001jointly,rizopoulos2011dynamic,kim2013joint,barrett2015joint,wu2014joint,chen2014joint}. In \eqref{bio-stat longi}, $a=(\alpha,\beta)$, $\alpha$ is a constant parameter vector, while $\beta$ is a random vector assumed to be normally distributed with mean $\mathcal{M}_1$ and co-variance $\Sigma_1$. $e$ is a white-noise observational error subjecting to mean $\mathcal{M}_2$ and co-variance $\Sigma_2$. $\mathcal{Z}_1$, $\mathcal{Z}_2$ are the so-called configuration matrices which are essentially two deterministic and known functions in variable $t$. Due to the existence of unobservable $\beta$, estimation of model \eqref{bio-stat longi} is based on Expectation-Maximization (EM) algorithm \citep{wu2007joint,rizopoulos2010jm}.

Longitudinal processes do not have to be continuous in all dimensions, it can take the jumped counting process as below:
\begin{equation}\label{count longi}
\mathcal{Z}\left(t,a\right)=\#\left\{\textrm{E}_{\tau}=1:\tau\in (0,t)\right\}
\end{equation}
where $\textrm{E}_{\tau}$ is the indicator to the occurrence of some recurrent events at time $\tau$ which is supposed to following some underlying distribution subjecting to parameter $a$, $\#$ operation counts the cardinality of the given set. Model \eqref{count longi} is a special case of panel count models \citep{riphahn2003incentive,sun2014panel}, many procedures are proposed for its estimation and a MPL-based estimation is discussed in \cite{andersen1982cox,andersen1992repeated}. 

Although estimation of model \eqref{joint model} with different longitudinal processes has been extensively discussed, in most cases, the discussion focus merely on that there are sufficient amount of longitudinal observations. Availability of longitudinal data may not be a big issue for clinical trails, but in many other applications, such as medical cost study and default risk forecast, it is often the case that longitudinal observations are systematically missed \citep{laird1988missing,hogan1997model,chen2014joint,sattar2017joint}. For instance of medical cost study, in order to protect privacy, many publicly available medical cost databases do not release longitudinal observations of the cost accumulation process during inpatient hospital stay, except the total charge by the discharge day. Medical databases of this type include the National Inpatient Sample (NIS) of United State and the New York State's Statewide Planning and Research Cooperative System (SPARCS). In the study of default rate of small business loans and/or consumer loans, all key financial indicators of borrowers are missing during the entire repayment period, this is because for small business and individual borrowers, financial data is even never collected unless some crucial event occurs as a trigger, such as overdue and/or default. So, it is common in financial applications that values of key variables are only available at the event time, all intermediate observations are missed.

To handle missing longitudinal data, a novel simulation-based estimation procedure is proposed in this paper. Without loss of generality, it is designed for the following form of input data, which is the typical data type in studies of default risk and medical cost:

\begin{equation}\label{partially missing data}
\left\{\left\{T_i,\left\{Z_{i,k,T_i}:k\in \mathcal{K}\right\},\left\{\left(Z_{i,k,t_{i,j}}\right)_{j=1}^{m_i}:k\in P/\mathcal{K}\right\}\right\}:i=1,\dots,n\right\}
\end{equation}
where $n$ subjects are in the input data, $m_i$ is the number of longitudinal observation times for $i$; $t_{i,j}$ and $Z_{i,k,t_{i,j}}$ are the $j$th observation time for subject $i$ and the value of the $k$th variable observed at time $t_{i,j}$ for $i$, respectively. For observation times, we assume the last observation time $t_{i,m_i}$ is always equal to the event time $T_i$ (censoring can be easily incorporated whenever it is uninformative, but it is not the major topic in this study). $P=\{1,\dots,\mathfrak{p}\}$ denotes the set of indices of $\mathfrak{p}$ longitudinal variables. There are two subsets of variables, $\mathcal{K}$ is the set of variables with their longitudinal observations systematically missed except for the observation at event time; while the complement set $P/\mathcal{K}$ consists of variables for which longitudinal observations are available at all observation time. 

The simulation-based procedure turns out capable of generating consistent estimators of all parametric and non-parametric components of model \eqref{joint model} from input data of type \eqref{partially missing data}. To our best knowledge, this is the first procedure that can handle \eqref{partially missing data}. Apart from feasibility, our simulation-based procedure is uniformly applicable to all three different specifications of longitudinal process \eqref{stationary cox}, \eqref{bio-stat longi} and \eqref{count longi}, as well as their mixture. The uniform property makes our procedure attractive as most existing procedures are either designed for continuous longitudinal process \eqref{bio-stat longi} \citep{andersen1982cox,karr2017point,lin2000semiparametric} or the counting process \eqref{count longi} \citep{taylor1994stochastic,kim2013joint,zeng2007maximum}, their combination is rarely discussed together with survival analysis. In addition, uniformity makes it possible to integrate estimation of different classes of joint models into one single software package, so it makes our procedure friendly to practitioners.

From the perspective of computation, the simulation-based procedure outperforms many existing procedures \citep{rizopoulos2010jm,guo2004separate} in the sense of being compatible with parallel computing framework. In fact, there are two major steps involved in the approach, a simulation step and a optimization step. The simulation step is carried out path-wisely so is completely parallelizable. From the simulation result, a complete version of likelihood function is derivable without any latent variable involved, so the optimization step doesn't rely on EM algorithm and can be parallelized as well if stochastic descending algorithm is applied. Its compatibility to parallel computing makes the simulation-based procedure useful in handling massive data and financial applications.

The paper is organized as the following. In Section 2, we will present model specification and sketch the estimation procedure in details.
The large sample properties of resulting estimators are stated in
Section 3. Simulation results and the application to massive consumer-loan data
are presented in Section 4. Section 5 discusses some extensions of
our model and concludes. All proofs are collected in Appendix.

\section{Model Specification \& Estimation}\label{model specification}
\subsection{Model Specification}\label{model_specification}
In this study, we consider a $\mathfrak{p}$-dimensional mixture longitudinal processes, its projection to the $i$th dimension, $Z_i$, is either a counting process of type \eqref{count longi} or absolutely continuous in time which, without loss of generality, can be expressed as a time integral:

\begin{equation}\label{time-integral longi}
Z_i\left(t\right)=Z_{i0}+\int_{0}^{t}\epsilon_i(s)ds
\end{equation}
where $\epsilon_i(t)$ is an arbitrary stochastic process with finite first and second moments at every $t$, $Z_{i0}$ is an arbitrary initial random variable. \eqref{time-integral longi} include both of \eqref{stationary cox} and \eqref{bio-stat longi} as special cases, it reduces to \eqref{stationary cox} as long as $\epsilon_i(t)\equiv 0$ and $Z_{i0}\equiv c$, and reduces to \eqref{bio-stat longi} if $\mathcal{Z}_1$ and $\mathcal{Z}_2$ in \eqref{bio-stat longi} are absolutely continuous with respect to $t$, which is usually assumed to hold in practice. 

To be general, it is allowed that some of the longitudinal dimension is not continuous in time, say representable as a counting process. For simplicity of presentation, we consider the case where $Z_i$ is a counting process only for one covariate dimension, while our methodology is easily extendible to the multi-dimensional counting process situation without much modification. So, from now on, we will denote longitudinal process as $Z(t)=(Z^{\ast}(t),Z^{-\ast}(t))$ with $Z^{\ast}$ and $Z^{-\ast}$ representing the counting dimension and absolutely continuous dimensions respectively.

For the counting dimension, following \cite{andersen1982cox,lin2000semiparametric}, we assume the conditional jumping intensity of the counting process is given through a cox model:

\begin{equation}\label{jump intensity}
\lambda^{c}\left(t,Z(t)\right)=\lambda^{c}_0(t)\exp\left(b^{c\top}Z(t)\right)
\end{equation}
where the supscript $c$ distinguish the intensity of the counting process from the intensity of the terminal event process.

\subsection{Simulatibility and simulation procedures}\label{simulate}
Since our estimation procedure is simulation-based, it requires the entire longitudinal process to be {\bf simulatible} which is formally defined as below:

\begin{definition}\label{simulatible}
A longitudinal process $Z(t)$ is simulatible if for every $dt>0$, it is possible to generate a sequence of random vectors $\{\zeta_i:\,i=0,1,2,\dots\}$ such that the process
\begin{equation}\label{approximation process}
Z'(t)=\zeta_{\left\lfloor t/dt\right\rfloor}
\end{equation}
converges weakly to the true process $Z(t)$ as $dt\rightarrow 0$, where $\lfloor t/dt\rfloor$ denotes the integral part of $t/dt$.

In addition, a longitudinal process $Z(t)$ is {\bf empirically simulatible} if there exists a simulation procedure such that for every positive integer $N$, $k$ and every $dt>0$, $N$ identically independently distributed (i.i.d.) sample sequences of the form \begin{equation}\label{empirical approximation process}
\{\zeta_{i,l}:\,l=1,\dots,k\}, \,i=1,\dots,N
\end{equation} can be generated from the procedure, such that for every $l\leq k$, the cross section $\{\zeta_{i,l}:\,i=1,\dots,N\}$ is $N$ i.i.d. samples of $\zeta_{l}$ with $\zeta_{l}$ being the $l$th element of the simulatible sequence in \eqref{approximation process} corresponding to $dt$. 
\end{definition} 

Most widely-studied longitudinal processes are empirically simulatible. Process of type \eqref{stationary cox} is simulatible through computing the time integral \eqref{time-integral longi} with $\epsilon\equiv 0$. Process \eqref{bio-stat longi} is simulatible by the following algorithm:

\begin{minipage}{\linewidth}
\begin{algorithm} [H]
\caption{{\bf GenSim}--generate simulatible sequence for \eqref{bio-stat longi}}\label{main algorithm 0}
\begin{algorithmic} [1] 
\Require \\
constant parameter, $\alpha$\\
distribution of $\beta$ in \eqref{bio-stat longi}, $F_{\beta}$\\ 
distribution of $\alpha\cdot\mathcal{Z}_1(0)+\beta\cdot\mathcal{Z}_2(0)+e$ in \eqref{bio-stat longi}, $F_{0}$\\
configuration matrices $\mathcal{Z}_1$ and $\mathcal{Z}_2$\\
initial time, $t$\\
 interval length, $dt$\\
sample size, $N$
\Ensure \\
A sequence of random variables satisfies \eqref{empirical approximation process}, denote as $V$.
 \algrule
 \State $\textrm{Set } V=\emptyset$
 \State $\textrm{Draw }N \textrm{ independent samples from distribution }F_{0} \textrm{, denote }\hat{Z} \textrm{ as the set of samples}$
 \State $V\varB{.append}(\hat{Z})$
 \State $\textrm{Draw }N \textrm{ independent samples from distribution }F_{\beta} \textrm{, denote }\hat{\beta} \textrm{ as the set of samples}$
\For {$\textbf{each } i \textbf{ in } \mathbb{N}$} 
	\State $\textrm{Set }\hat{Z}=V\varB{.last}$
	\State $\textrm{Set }\hat{Z}_1=\emptyset$
	\For {$ j=1 \textbf{ to } N$}
		\State $\textrm{Set }z=\hat{Z}[j]$ 
		\State $\textrm{Set }\beta=\hat{\beta}[j]$
		\State $\textrm{ReSet }z+=\alpha\cdot (\mathcal{Z}_1(t[j]+i\cdot dt)-\mathcal{Z}_1(t[j]+(i-1)\cdot dt))+\beta\cdot (\mathcal{Z}_2(t[j]+i\cdot dt)-\mathcal{Z}_2(t[j]+(i-1)\cdot dt))$ 
		\State $\hat{Z}_1\varB{.append}\left(z\right)$
	\EndFor
	\State $V\varB{.append}(\hat{Z}_1)$
	\EndFor\\
\Return $V$    
\end{algorithmic}
\end{algorithm}
\end{minipage}

\vspace{0.3cm}
In algorithm \ref{main algorithm 0}, $V\varB{.last}$ is the last element of the sequence $V$, $x[j]$ denotes the $j$th element in a sequence $x$; $\mathbb{N}$ is the set of natural number starting from $1$; $append$ operation appends an element to the given sequence. Note that initial time $t$ is a $N$-dimensional vector and considered as an input, whose entries are constantly $0$ in most cases. But algorithm \ref{main algorithm 0} and the following algorithm \ref{main algorithm 1} can be used as an intermediate step in the simulation of longitudinal process with counting component. In that case, initial time $t$ may not always be zero, so we leave $t$ as a free parameter.

In general, denote $\epsilon(t)$ as the vector $(\epsilon_1(t),\dots,\epsilon_{\mathfrak{p}}(t))$, with $\epsilon_i$ the variational rate of the $i$th-dimension time integral in \eqref{time-integral longi}, when $\epsilon(t)$ is Markovian in the following sense: 
\begin{equation}\label{markovian}
p\left(\epsilon(t)\in x+dx\mid Z(s),\,s\leq t\right)=p\left(\epsilon(t)\in x+dx\mid Z(t),t\right),
\end{equation}
absolutely continuous process \eqref{time-integral longi} is simulatible via the following Euler-type algorithm, where $p(\cdot\mid Z(s),\,s\leq t)$ is the conditional density of $\epsilon(t)$ given the full information of longitudinal process up to time $t$, $p(\cdot\mid Z(t),t)$ is the conditional density given only the longitudinal observation at time $t$:

\begin{minipage}{\linewidth}
\begin{algorithm} [H]
\caption{{\bf GenSim1}--generate simulatible sequence for \eqref{time-integral longi}}\label{main algorithm 1}
\begin{algorithmic} [1] 
\Require \\ 
longitudinal parameter, $a$\\
Distribution of $Z_0$ for fixed $a$, $F_{0}(\cdot,a)$, or a set of $N$ samples of $Z_0$, $S_N$\\ 
conditional pdf of the form \eqref{markovian}, $p$\\
initial time, $t$\\
 interval length, $dt$\\
 sample size, $N$
\Ensure \\
A sequence of random variables satisfies \eqref{empirical approximation process}, denote as $V$.
 \algrule
 \State $\textrm{Set } V=\emptyset$
 \State $\textrm{Set }\hat{Z}=S_N \textrm{ or the set of }N \textrm{ independent samples drawn from distribution }F_{0}(\cdot,a)$
 \State $V\varB{.append}(\hat{Z})$
\For {$\textbf{each } i \textbf{ in } \mathbb{N}$} 
	\State $\textrm{Set }\hat{Z}=V\varB{.last}$
	\State $\textrm{Set }\hat{Z}_1=\emptyset$
	\For {$j=1\textbf{ to } N$} 
		\State $\textrm{Draw a random sample, }z' \textrm{, from conditional density }p(\cdot \mid z,t)$
		\State $\hat{Z}_1\varB{.append}(z+z'\cdot dt)$
	\EndFor
	\State $V\varB{.append}(\hat{Z}_1)$
	\EndFor\\
\Return $V$    
\end{algorithmic}
\end{algorithm}
\end{minipage}\\

Longitudinal process with one counting dimension is also empirically simulatible, while the simulation algorithm becomes tricky. For simplicity of presentation, in the following pseudo-code (algorithm \ref{main algorithm 2}), we assume that the first dimension (indexed by $0$) of the longitudinal process represents the counting component. In addition, we require that at initial time, the counting dimension puts all mass at $0$, this restriction can be easily relaxed but the pseudo-code would become too redundant.

\begin{algorithm}
\small
\caption{{\bf GenSim2}--generate simulatible sequence with one counting dimension}\label{main algorithm 2}
\begin{algorithmic} [1] 
\Require \\
longitudinal setup, $\Omega=(a, b^c,\lambda_0^c)$\\
distribution of $Z_0$ for fixed longitudinal setup, $F_{0}(0,\cdot,\Omega)$; or a set of $N$ samples of $Z_0$, $S_N$\\ 
conditional pdf of form \eqref{markovian}, $p$\\
 interval length, $dt$\\
 sample size, $N$ 
\Ensure \\
A sequence of random variables satisfies \eqref{empirical approximation process}, denote as $V$.
 \algrule
 \State $\textrm{Set }\hat{Z}=S_N \textrm{ or the set of }N \textrm{ independent samples drawn from distribution }F_{0}(0,\cdot,\Omega)$
\State $\textrm{Set }V=\emptyset$
\State $\textrm{Set }jump\_t \textrm{ as an }N\textrm{ dimensional vector with all entries being }0$
\For {$\textbf{each } i \textbf{ in } \mathbb{N}$} 
\State $\textrm{ReSet }p(j,x\mid z,t)=\begin{cases}
p(x\mid z,t)&\textrm{ if }j=z^{\ast}\\
0 &\textrm{ else}
\end{cases}$
	\State $\textrm{Set }V' = \varB{GenSim1}(\Omega,\hat{Z},p,jump\_t,dt,N)$ 
	\If {$i=1$}
	\State $\textrm{ReSet }V=V'$
	\Else
	\For {$j=1 \textrm{ to }N \textrm{ and } k\geq jump\_t[j]$}
	\State $\textrm{ReSet }V[k][j]=V'[k-jump\_t[j]][j]$
	\EndFor
	\EndIf	
	\For {$\textbf{each } k \textbf{ in } \mathbb{N}$}
	\State $\textrm{ReSet }V'[k]=\sum_{j=1}^k \exp\left( - \exp\left(\varB{dot}(V'[k]\cdot dt,b^c\right)\cdot \lambda_0^c(k\cdot dt)\right)$
	\EndFor

	\For {$ j=1 \textbf{ to } N$}
	\State $\textrm{Set }\omega=uni(0,1)$
	\State $\textrm{Set }k=\min\{k'\in \mathbb{N}:\,V'[k'][j]\leq \omega\}$
	\State $\textrm{ReSet }jump\_t[j]=k$

	\For {$k'\geq k$}
	\State $\textrm{ReSet }V[k'][j][0]+=1$
	\EndFor 

	\State $\textrm{ReSet }\hat{Z}[j]=V[jump\_t[j]][j]$
	\EndFor
	\EndFor\\
	
\Return $V$    
\end{algorithmic}
\end{algorithm}
In algorithm \ref{main algorithm 2}, for every $k\in\mathbb{N}$, $V[k]$ is considered as $N\times \mathfrak{p}$ matrix with its $j$th row $V[k][j]$ being the $\mathfrak{p}$ dimensional longitudinal process simulated at time $k\cdot dt$ and $V[k][j][0]$ denote the corresponding value at the counting dimension. $\varB{dot}(.,.)$ is the inner product between matrices with appropriate row- and column-dimensions, the product matrix $V'[k]\cdot dt$ is formed through entry-wise product between $dt$ and entries in $V'[k]$. $uni(0,1)$ is a random number generator that draws a random number from uniform distribution on interval $[0,1]$.

It is critical to notice that the counting component can be considered as time-invariant between two consecutive jump time (reflected through the line 11 of algorithm \ref{main algorithm 2}, where the conditional density is reconstructed and the probability that the counting component jumps out of its current stage $z^{\ast}$ is set to $0$), then the design of algorithm \ref{main algorithm 2} becomes quite simple. 
It takes fully use of the local stationarity of the counting component in the way that for every fixed subject, simulation of the longitudinal process with one counting dimension is decomposed as a sequential simulation of longitudinal processes with all their dimensions being absolutely continuous in time. This sequential construction is crucial not only in algorithm design, but is also the key to verify the identifiability of model \eqref{joint model} under specification \eqref{time-integral longi} and \eqref{jump intensity}, the details will be discussed in section \ref{large sample property all}.

It turns out that algorithm \ref{main algorithm 0}-\ref{main algorithm 2} can generate the simulatible sequence required in definition \ref{simulatible} for all longitudinal processes \eqref{stationary cox}, \eqref{bio-stat longi}, \eqref{count longi} and their mixture. Proof for algorithm \ref{main algorithm 0} is quite trivial, while proof for algorithm \ref{main algorithm 1} and \ref{main algorithm 2} are parts of the proof of the consistency of our estimators which, therefore, are combined with the proof of theorem \ref{theorem consistency} and presented in Appendix. 

Notice that algorithm \ref{main algorithm 2} is extendible to handle the occurrence of the terminal event. In fact, algorithm \ref{main algorithm 2} can be generalized to the following algorithm \ref{main algorithm 3}, which returns a set of i.i.d. samples of $(Z_T,T)$. $(Z_T,T)$ is the joint of longitudinal observation at event time and the event time itself, its i.i.d. samples are the key to construct the joint probability density function (pdf) and likelihood function in our estimation procedure.

\begin{algorithm} 
\caption{{\bf GenSim3}--generate joint samples of longitudinal and event time data }\label{main algorithm 3}
\begin{algorithmic} [1] 
\Require \\
longitudinal setup, $\Omega=(a, b,b^c,\lambda,\lambda^c)$\\
distribution of $Z_0$ for fixed longitudinal setup, $F_{0}(0,\cdot,\Omega)$; or a set of $N$ samples of $Z_0$, $S_N$\\ 
conditional pdf of the form \eqref{markovian}, $p$\\
initial time, $t$\\
 interval length, $dt$\\
 sample size, $N$ \\
 censor bound, $C$
\Ensure \\
$N$ samples of the pair $(Z_{T},T)$ of longitudinal variables at event time.
 \algrule
\State $\textrm{Set }V=\emptyset$
 \State $\textrm{Set }\hat{Z}=S_N \textrm{ or the set of }N \textrm{ independent samples drawn from distribution }F_{0}(0,\cdot,\Omega)$
\State $\textrm{Set }jump\_t \textrm{ as an }N\textrm{ dimensional vector with all entries to be }0$
\State $\textrm{Set }event\_t \textrm{ as an }N\textrm{ dimensional vector with all entries to be }C$
\While {$\{t:\,t\in event\_t,\,t<C\}\not=event\_t \textrm{ and } \{t:\,t\in jump\_t,\,t\geq C\}\not=\emptyset$} 
\State $\textrm{ReSet }p(j,x\mid z,t)=\begin{cases}
p(x\mid z,t)&\textrm{ if }j=z^{\ast}\\
0 &\textrm{ else}
\end{cases}$
	\State $\textrm{Set }V' = \varB{GenSim1}(\Omega,\hat{Z},p,jump\_t,dt,N)$ 
	\State $\textrm{Set }V^{''} = V'$
	\If {$i=1$}
	\State $\textrm{ReSet }V=V'$
	\Else
	\For {$j=1 \textrm{ to }N \textrm{ and } k\geq jump\_t[j]$}
	\State $\textrm{ReSet }V[k][j]=V'[k-jump\_t[j]][j]$
	\EndFor
	\EndIf	
	\For {$\textbf{each } k \textbf{ in } \mathbb{N}$}
	\State $\textrm{ReSet }V^{''}[k]=\sum_{j=1}^k \exp\left( - \exp\left(\varB{dot}(V'[k]\cdot dt,b\right)\cdot \lambda(k\cdot dt)\right)$
	
	\State $\textrm{ReSet }V'[k]=\sum_{j=1}^k \exp\left( - \exp\left(\varB{dot}(V'[k]\cdot dt,b^c\right)\cdot \lambda^c(k\cdot dt)\right)$
	\EndFor
	
	\For {$ j=1 \textbf{ to } N$}
	\State $\textrm{Set }\omega=uni(0,1)$
	\State $\textrm{Set }\omega'=uni(0,1)$
	\State $\textrm{Set }k=\min\{k'\in \mathbb{N}:\,V'[k'][j]\leq \omega\}$
	\State $\textrm{Set }k^{\ast}=\min\{k'\in \mathbb{N}:\,V^{''}[k'][j]\leq \omega'\}$
	\State $\textrm{ReSet }jump\_t[j]=k$
	\If {$k^{\ast}<k$}
	\State $\textrm{ReSet }event\_t[j]=k^{\ast}$
	\EndIf
\algstore{myalg}
\end{algorithmic}
\end{algorithm}

\begin{algorithm}                     
\begin{algorithmic} [1]                   
\algrestore{myalg}
	\For {$k'\geq k$}
	\State $\textrm{ReSet }V[k'][j][0]+=1$
	\EndFor 
	\State $\textrm{ReSet }\hat{Z}[j]=V[jump\_t[j]][j]$
	\EndFor
\EndWhile
\State $\textrm{Set }Sample=\emptyset$
\For {$i=1 \textrm{ to }N$}
\State $k=event\_t[i]$
\State $Sample\varB{.append}(\varB{concat(V[k][i],\{k\cdot dt\})})$
\EndFor\\
\Return $Sample$    
\end{algorithmic}
\end{algorithm}
%
In algorithm \ref{main algorithm 3}, all notations follow their interpretations in previous three algorithms. Censor bound $C$ is a prescribed positive constant, it specifies the end of observation. $concat$ operation returns a $n1+n2$ dimensional vector through concatenating two row vectors with dimension $n1$ and $n2$, respectively. Apparently, algorithm \ref{main algorithm 3} simulates the terminal event time in the same way as to simulate the jump time of counting component in algorithm \ref{main algorithm 2}.

\subsection{Estimation procedure}\label{estimation}

The simulation algorithms stated in previous section provide a foundation to construct the estimation procedure of model \eqref{joint model}. Our estimation is based on maximizing the full information likelihood function of observations $\{(Z_{T_i},T_i):\,i=1,\dots,n\}$. 

Notice that at this moment, we assume that longitudinal observations are systematically missed for all covariate dimensions, i.e. the input data has a special form of \eqref{partially missing data} with the set $\mathcal{K}=P$. Estimation for more general form of input data is extended through the estimation procedure of the special case. To avoid ambiguity of notation, from now on, we will denote $S$ as the terminal event time derived from simulation and $W_t$, $W_S$ as the longitudinal observation of simulated sample at $t$ and terminal time $S$ respectively. In contrast, for the real observed sample, the terminal event time is denoted as $T$ and longitudinal observations are denoted as $Z_t$ or $Z_T$.

With the aid of simulation algorithms presented in previous section, the construction of likelihood function can be implemented according to following two steps:

\noindent {\bf Step 1}: Fix a sample size $N$, interval length $dt$, and a profile of model parameters $(a,b,b^c)$ and non-parametric components $(\lambda_0,\lambda_0^c)$, executing appropriate simulation algorithms yields $N$ samples of $\left\{(W_{S_i},S_i):\,i=1,\dots,N\right\}$ which turn out to be i.i.d. samples of $(W_S,S)$ subjecting to the given parameters. \\
\noindent {\bf Step 2}: Apply kernel density method to the i.i.d. samples in Step 1, yield an empirical pdf of the random vector $(W_S,S)$ that implicitly depends on $(a,b,b^c,\lambda_0,\lambda_0^c)$ and is expressed as below: 
\begin{equation}\label{empirical pdf}
\hat{p}_{N,dt,h}(z,s |a,b,b^c,\lambda_0,\lambda_0^c)=\frac{1}{N}\sum_{i=1}^NK^{(\mathfrak{p}+1)}_h\left(z-W_{S_i},s-S_i\right)
\end{equation}
where $K_h^{(m)}$ is a $m$-dimensional kernel function with bandwidth $h$, for simplicity, we only consider the Gaussian kernel function in this study; $(W_{S_i},S_i)$ are samples yielding from step 1.

The i.i.d. property of samples from Step 1  guarantees that $\hat{p}_{N,dt,h}(\cdot |\Omega)$ yielding from Step 2 converges to the true joint pdf subjecting to $\Omega=(a,b,b^c,\lambda_0,\lambda_0^c)$.

Based on empirical pdf $\hat{p}_{N,dt,h}(\cdot |\Omega)$, an empirical version of the full information likelihood function can be constructed as below:

\begin{equation}\label{likelihood}
l_{n,N,dt,h}(a,b,b^c,\lambda_0,\lambda_0^c)=\prod_{i=1}^{n}\hat{p}_N(Z_{T_i},T_i |a,b,b^c,\lambda_0,\lambda_0^c)
\end{equation}
where $(Z_{T_i},T_i)$ is the observed covariate variables at the event time $T_i$ of the $i$th subject and the event time itself, $n$ is the sample size (the number of subjects) of the input data.

In practice, the non-parametric components $\lambda_0$ and $\lambda_0^c$ in \eqref{likelihood} can be replaced by their step-wise version:
\begin{equation}\label{step-wise-lambda}
\lambda\left(t\right):=\sum_{i=1}^{k}\theta_{i}\cdot I\left(t\in[dt\cdot(i-1),dt\cdot i)\right)
\end{equation}
where $\lambda$ in \eqref{step-wise-lambda} can take either as $\lambda_0$ or $\lambda_0^c$, $k=\min\left\{k'\in \mathbb{N}: k\cdot dt>\max\{T_i:\,i=1,\dots,n\}\right\}$, $I$ is the indicator function, for simplicity, the step length $dt$ is taken the same value as the length of time interval in simulation algorithm \ref{main algorithm 0}-\ref{main algorithm 3}. To guarantee consistency, their values will depend on the sample size $n$.

Substitute $\lambda_0$ and $\lambda_0^c$ of the form \eqref{step-wise-lambda} into \eqref{likelihood} yielding the final form of likelihood function. Our estimator $(\hat{a},\hat{b},\hat{b}^c,\hat{\lambda}_0,\hat{\lambda}_0^c)$ is then derived through maximizing the renewed version of likelihood function \eqref{likelihood}.

\subsection{Estimation with general input-data type \eqref{partially missing data}}\label{estimation_extension}
Although the estimation procedure of maximizing \eqref{likelihood} is directly applicable to the input data of type \eqref{partially missing data}, it does not fully utilize the information provided in input data as there is not any connection between the estimator and the partially existed longitudinal observations. To resolve this issue, we present a way to extend the estimation procedure in previous section, the extension can make better use of longitudinal information and increase estimation efficiency.

We apply the idea of censoring to construct a weighted average likelihood function, estimators fully utilizing the longitudinal data is then derived from maximizing that function. In details, for every fixed time interval $[t,t')$ and the subset of simulated data satisfying $S_i\geq t$, the simulated samples admit to construct the conditional joint pdf of $(W_S,S)$ given $t\leq S< t'$ and the conditional density of censored event $W_t=z$ given the censoring $S>t$. More precisely, given $t$ and $t'>t$, we have the following uncensored pdf:

\begin{equation}\label{uncensored pdf}
\hat{p}_{N_t,dt,h}^{t,t',u}\left(z,s\mid a,b,b^c,\lambda_0,\lambda_0^c\right)=\left(\frac{1}{N_t}\sum_{i=1}^NK^{(\mathfrak{p}+1)}_h\left(z-W_{S_i},s-S_i\right)\cdot I(t\leq S_i\leq t')\right)^{I(t\leq s<t')};
\end{equation}
and the censored pdf

\begin{equation}\label{censored pdf}
\hat{p}_{N_t,dt,h}^{t',c}\left(z^{\mathcal{K}}\mid a,b,b^c,\lambda_0,\lambda_0^c\right)=\left(\frac{1}{N_t}\sum_{i=1}^NK^{(|\mathcal{K}|)}_h\left(z^{\mathcal{K}}-W_{S_i}^{\mathcal{K}}\right)\cdot I(S_i> t')\right)^{I(s\geq t')}
\end{equation}
where $K_h^{(.)}$ follows the interpretation before; $N_t<N$ is the number of simulated subject $i$s with terminal event time $S_i\geq t$; $|\mathcal{K}|$ is the number of element in the set of dimensions that do not have missing longitudinal observations; $x^{\mathcal{K}}$ denotes the projection of vector $x$ onto its sub-coordinate indexed by $\mathcal{K}$; supscript $t$, $t'$ indicates the dependence on $t$ and $t'$, supscript $u$ and $c$ denote \enquote{uncensored} and \enquote{censored} respectively.

With the aid of \eqref{uncensored pdf} and \eqref{censored pdf}, we can partition the time line $[0,\max\left\{T_i:\,i=1,\dots,n\right\}]$ into $m$ disjoint intervals with their boundaries recorded as $0=t_0<t_1<\dots<t_m$, then the mean log-likelihood function is constructed as below:

\begin{equation}\label{mean likelihood}
\Scale[0.8]{
l_{n,N,dt,h,m}(\Omega)=\frac{1}{m}\sum_{j=1}^{m}\frac{1}{n_{t_{j-1}}}\sum_{i=1}^n\left(\log\hat{p}_
{N_{t_{j-1}},dt,h}^{t_{j-1},t_j,u}\left(Z_{i,T_{i}},T_i\mid \Omega\right)+\log\hat{p}_{n_{t_{j-1}},dt,h}
^{t_{j-1},c}\left(Z_{i,\mathcal{K},t_{i,j}}\mid \Omega\right)\right)}
\end{equation}
where $Z_{i,t}$ is the real observed longitudinal vector for subject $i$ at time $t$; $Z_{i,\mathcal{K},t_{i,j}}$ is the observed sub-coordinates of longitudinal vector of subject $i$ at time $t_{i,j}$ corresponding to those non-missing dimensions $\mathcal{K}$; $\Omega=(a,b,b^c,\lambda_0,\lambda_0^c)$; $n_t$ is defined analogous to $N_t$ by replacing simulated sample with real observed sample, formally, $n_t=\#\{i\in\{1,\dots,n\}:T_i\geq t\}$.

Estimators fully utilizing longitudinal information is derived from maximizing the mean likelihood function \eqref{mean likelihood} and denoted as $(\hat{a}^l,\hat{b}^l,\hat{b}^{c,l},\hat{\lambda}^l_0,\hat{\lambda}_0^{c,l})$.

\begin{remark}\label{remark1}
The choice of partition boundaries of $\{t_1,\dots,t_m\}$ and the number of partition cells $m$ is tricky and input-dependent. 

When the observation time in \eqref{partially missing data} is uniform for all subject in sample, i.e. $t_{i,j}\equiv t_{i',j}$ for all different $i$, $i'$ and $j<\min(m_i,m_{i'})$, partition boundaries of $\{t_1,\dots,t_m\}$ can be simply selected as $\{t_{i^{\ast},1},\dots,t_{i^{\ast},m_{i^{\ast}}}\}$ where $i^{\ast}$ is the index of the subject who has the greatest terminal event time. This choice can guarantee the most efficient utilizing longitudinal information. Input data with uniform observation time is widely existing in applications to finance and actuarial sciences where many economic variables are collected in a fixed frequency, such as GDP \citep{koopman2008multi,li2017impact}.

When the observation time is not uniform, but for every subject in sample, the observation frequency is relatively high in the sense that $\Delta=\max\{t_{i,j+1}-t_{i,j}:\,i=1,\dots,n;\,j=1,\dots,m_i\}$ converges to $0$ as sample size $n\rightarrow \infty$, partition intervals can be selected with equal length while the number of partition intervals is set as $n$. In this case, interpolation method can be applied as discussed in \cite{andersen1982cox} to set longitudinal value at boundary point $t_j$ for subjects whose longitudinal observation are missing at $t_j$.

Finally if longitudinal observation time is not uniform and has low frequency, the choice of $\{t_1,\dots,t_m\}$ and $m$ becomes quite complicated, we leave it for future discussion.
\end{remark}


\subsection{Parallel computing}\label{parallel computing}
Distinct from the estimators for joint models with longitudinal process specified through \eqref{bio-stat longi} \citep{wu2014joint,rizopoulos2010jm,guo2004separate}, the computation of our estimator is highly compatible with parallel computing framework, especially with the embarrassingly parallel computing framework \cite{guo2012parallel}. The parallelizability of our procedure comes from two sources which correspond to the simulation steps and optimization steps, respectively. 

In the step of simulation and construction of empirical pdf, all simulation algorithms in section \ref{model_specification} is implemented in a path-wise manner, In fact, setting sample size $N=1$ for each run of algorithm \ref{main algorithm 0}-\ref{main algorithm 3} and repeating execution for $N>1$ times generates $N$ samples that are essentially identical to the samples by executing the algorithms once with sample size $N$. So there are no interaction between two sample trajectories, an embarrassingly parallel computing framework is perfectly applicable in this setting and can significantly rise up the computation speed of the simulation step.

In the step of optimization, there is no latent variable involved in the empirical likelihood function \eqref{likelihood}, this is quite different from the estimation procedure of joint models with \eqref{bio-stat longi} as longitudinal process, where the involvement of random coefficient $\beta$ leads to latent variables and the reliance on EM algorithm. The main advantage of not using EM algorithm is that there is no need to repeatedly solve a complex integral and a maximization problem which have sequential dependence.  Consequently, evolutionary algorithm and/or stochastic descending algorithm \citep{liu2015asynchronous,cauwenberghs1993fast,sudholt2015parallel,tomassini1999parallel,osmera2003parallel} is applicable to maximize the likelihood function \eqref{likelihood}, which is embarrassingly parallelizable.

In sum, the estimation proposed in this paper is highly compatible with parallel computing framework. This is important because in applications to finance or risk management, massive input data is common that imposes a strict requirement on computation speed and efficient memory allocation. Parallel computing can significantly lift computation speed, meanwhile take better use of the limit memory. So, being parallelizable grants our estimation procedure with a great potential in a wide range of applications, especially the application in finance.
  
\section{Asymptotic Property of Large Sample}\label{large sample property all}\label{asymptotic property}

The consistency and asymptotic normality of estimators, $(\hat{a},\hat{b},\hat{b}^c,\hat{\lambda}_0,\hat{\lambda}_0^c)$ are established
in this section. For convenience of expression, we need the following notations:

\noindent 1. Denote $A$, $B$ and $B^c$ as the domain of parameter $a$, $b$ and $b^c$, respectively.

\noindent 2. Denote $\Omega=(a,b,b^c,\lambda_0,\lambda_0^c)$ as a given model setup, with $\Omega_0$ being the true model setup; denote function $p$ as the theoretical joint pdf depending on model setup $\Omega$, $p(\cdot|\Omega_0)$ is the true joint pdf of observation $(Z_T,T)$.

\noindent 3. Denote 
\begin{equation}\label{function-q}
q(j,z,t|\Omega)=\textrm{E}\left(\epsilon(t)\mid Z^{\ast}(t)=j,Z^{-\ast}(t)=z,\Omega\right)
\end{equation}
as the conditional expectation of variational rate of absolutely continuous dimensions of longitudinal process given its observation at time $t$ and model setup $\Omega$, where $\epsilon(t)$ is the vector of instantaneous variational rate as defined in \eqref{time-integral longi}, $Z^{\ast}$ and $Z^{-\ast}$ denote the counting dimension and absolutely continuous dimension of longitudinal process respectively.
To establish the consistency result, we need the following technical conditions:

\noindent $\mathbf{C1}$. For every $j\in\mathbb{N}$, the pdf $f_0(j,.,\Omega_0)$ induced by the true longitudinal process at initial time has full support and satisfies that $\sum_{j\in\mathbb{N}}\int_{\mathbb{R}^{\mathfrak{p}-1}}f_0(j,z,\Omega_0)\cdot\exp(v\cdot z)dz\not=1$ for all non-zero $\mathfrak{p}-1$ dimensional vector $v$. In addition, $\lambda_0(0)\equiv C_1$, $\lambda_0^c(0)\equiv C_2$ for some positive constant $C_1,\,C_2$ and for all $\lambda_0,\,\lambda_0^c$ in consideration.

\noindent $\mathbf{C2}$. For every $\Omega\not=\Omega_0$ and every $j\in \mathbb{N}$, one of the following holds: 
\begin{itemize}
\item[(i)] $f_0(j,z,\Omega)\not\equiv f_0(j,z,\Omega_0)$;
\item[(ii)] There exists $t>0$ such that the matrix $\nabla_z q\left(j,z,t\mid \Omega\right)-\nabla_z q\left(j,z,t\mid \Omega'\right) $ converges to some limiting matrix as $\Vert z\Vert\rightarrow \infty$, denote the limiting matrix as $M$ which satisfies the hyperbolic property, i.e. at least one eigenvalue of $M$ must have non-zero real part.

\end{itemize}

\noindent $\mathbf{C3}$. $A$, $B$ and $B^c$ are compact subsets of Euclidean space with appropriate dimension, suppose that they all have open interiors and the true values of parameter vector $a$, $b$ and $b^c$ contained in their open interior.

\noindent $\mathbf{C4}$. $\textrm{E}_{\Omega_{0}}\left(\left|\log\left(p(Z_T,T |\Omega)\right)\right|\right)$,
$\textrm{E}_{\Omega_{0}}\left(\left|\nabla_{x}\log\left(p(Z_T,T |\Omega)\right)\nabla_{x'}\log\left(p(Z_T,T |\Omega)\right)\right|\right)$ and 
$\textrm{E}_{\Omega_{0}}(|\nabla^2_{xx'}\\
\log(p(Z_T,T |\Omega))|)$
are finite for all $x,\,x'$ as pairs of coordinates of vector $(a,b,b^c)$, and for all $\Omega$ in its domain; denote $\mathcal{I}$ as a matrix with its $xx'$ entry being $\mathcal{I}_{xx'}=\textrm{E}_{\Omega_{0}}(\nabla_{x}\log(p(Z_T,T |\Omega_0))\nabla_{x'}\log\\
(p(Z_T,T |\Omega_0)))$, denote $\mathcal{H}$ as a matrix with its $xx'$ entry being $\mathcal{H}_{xx'}= \textrm{E}_{\Omega_{0}}\left(\nabla^2_{xx'}\log\left(p(Z_T,T |\Omega_0)\right)\right)$, both matrices $\mathcal{I}$ and $\mathcal{H}$
are positive definite.

\noindent $\mathbf{C5}$. For all combinations of $\Omega$ and all $j\in \mathbb{N}$, $q\left(j,.\mid \Omega\right)\in C^{2}\left(\mathbb{R}^{\mathfrak{p}}\times\mathbb{R}_{+}\right)$;
for every $j\in \mathbb{N}$ the map from the domain of $\Omega$ to $C^{2}\left(\mathbb{R}^{\mathfrak{p}}\times\mathbb{R}_{+}\right)$ given through $\Omega\mapsto q\left(j,.\mid \Omega\right)$
is continuous with respect to the $C^{2}$ topology. 

\noindent $\mathbf{C6}$. The true theoretical joint pdf $p(\cdot |\Omega_0)$ is continuously differentiable with bounded first order partial derivatives.

\noindent $\mathbf{C7}$. $n$ is the number of subjects in observation. The choice of parameter $dt$, $N$ and kernel width $h$ satisfies $N \sim O(n)$, $dt\sim O(n^{-1})$, $nh^3\rightarrow 0$ and $nh\rightarrow \infty$.

Condition $\mathbf{C1}$ and $\mathbf{C2}$ are the key to verify model identification. Condition $\mathbf{C3}$-$\mathbf{C5}$ are the standard assumption in maximum likelihood estimation (MLE), which guarantees the consistency and asymptotic normality of MLE estimators. Condition $\mathbf{C6}$ and $\mathbf{C7}$ guarantee that the simulation algorithms introduced in section \ref{model_specification} can generate the required simulatible sequence in \ref{simulatible} and that the empirical joint pdf, $\hat{p}$ in \eqref{empirical pdf}, is consistent.

\begin{remark}
Among the seven conditions, $\mathbf{C1}$ and $\mathbf{C2}$ plays the central role to guarantee identifiability of our estimator. It turns out that $\mathbf{C1}$ and $\mathbf{C2}$ hold for a very general class of longitudinal processes. Particularly, almost all linear mixed longitudinal processes \eqref{bio-stat longi} are belonging to that class. In fact, for all absolutely continuous $\mathcal{Z}_1$, $\mathcal{Z}_2$ such that $\mathcal{Z}_2(0)\not=0$, longitudinal processes \eqref{bio-stat longi} can be rewritten as a time-integral of the form \eqref{time-integral longi}:
\begin{equation}
Z(t)=e+\alpha\cdot\mathcal{Z}_1(0)+\beta\cdot\mathcal{Z}_2(0)
+\int_0^t\alpha\cdot\mathcal{Z}_1'(s)+\beta\cdot\mathcal{Z}_2'(s)ds
\end{equation}
where the initial vector $Z_0=e+\alpha\cdot\mathcal{Z}_1(0)+\beta\cdot\mathcal{Z}_2(0)$
which, by assumption, is a normal random vector, so always satisfies $\mathbf{C1}$ and $\mathbf{C2}$ (i). In fact, $\mathbf{C1}$ and $\mathbf{C2}$ (i) does not only hold for the normal class, but also hold for most popular distribution classes that we can meet in practice. This distribution-free property is crucial, as the computation of the traditional estimators to model \eqref{joint model} is expensive in time and memory, it strongly relies on the normality assumption to simplify the expression of likelihood function \citep{rizopoulos2010jm}. However, computation load of our simulation-based estimator is not sensitive to the normality assumption, because according to algorithm \ref{main algorithm 0}, the variation of distribution class of $e$ and $\beta$ only affects the draw of initial samples, which does not take more time and/or memories for most of the widely-used distribution classes.

\end{remark}

\begin{remark}
$\mathbf{C1}$ and $\mathbf{C2}$ distinguish our simulation-based estimator from the traditional estimators developed for model \eqref{joint model} under longitudinal specification \eqref{bio-stat longi} \citep{rizopoulos2010jm,guo2004separate}. In literature, the standard trick is to consider the latent factor $\beta$ as a random effect, then model \eqref{joint model} becomes a frailty model, conditional independent assumption is utilized to derive explicit expression of the likelihood function and EM algorithm is applied to carry out the estimation. However, once if \eqref{joint model} is treated as a frailty model, its identifiability strongly depends on the number of longitudinal observations, which have to be greater than the dimension of longitudinal processes for the normally distributed $\beta$ as proved in \cite{kim2013joint}. Proof in \cite{kim2013joint} also relies on the assumption of normality, it is not clear if the same trick applies to more general distribution classes. The dependence on normality and the availability of enough amount of longitudinal observations restrict the usefulness of joint model \eqref{joint model} in many fields, such as the credit risk management and actuarial science \citep{li2017impact,koopman2008multi}, where longitudinal processes are not normal in general. More critically, in most cases the observation of covariate variables is only available for a couple of years and collected on a monthly or quarterly base, so only tens of longitudinal records are present. In contrast, there are usually ultra-high dimensional (e.g. hundreds) covariates present. Thus, identifiability of model \eqref{joint model} is always a big concern. For our procedure, condition $\mathbf{C1}$ and $\mathbf{C2}$ guarantee model identifiability without any extra restriction on the number of longitudinal observations, neither on the distribution class of $\beta$. In this sense our simulation-based procedure generalizes the standard estimation procedure of model \eqref{joint model}.
\end{remark}
\begin{theorem}\label{theorem consistency}
Model \eqref{joint model} is identifiable under condition $\mathbf{C1}$ and $\mathbf{C2}$,  where its longitudinal process is a mixture of absolutely continuous processes specified through \eqref{time-integral longi} and an one-dimensional counting process satisfying \eqref{jump intensity}. Additionally, if $\mathbf{C3}$-$\mathbf{C7}$ hold,
the estimator $(\hat{a},\hat{b},\hat{b}^c,\hat{\lambda},\hat{\lambda}_0^c)$ is consistent, asymptotically normally distributed for its parametric part $(\hat{a},\hat{b},\hat{b}^c)$, and has the following asymptotic property for its non-parametric part:\\
$ $

\noindent For the two baseline hazard functions $\lambda_0$ and $\lambda^c_{0}$, the estimator
$\hat{\lambda}_0$ and $\hat{\lambda}_0^c$ converge to $\lambda_{0}$ and $\lambda_0^c$ according to the
weak-$*$ topology, the processes $\sqrt{n}\left(\int_{0}^{t}\hat{\lambda}_0\left(\tau\right)-\lambda_{0}\left(\tau\right)d\tau\right)$ and $\sqrt{n}\left(\int_{0}^{t}\hat{\lambda}_0^c\left(\tau\right)-\lambda_{0}^c\left(\tau\right)d\tau\right)$
converge weakly to two Gaussian Processes.

\end{theorem}

The estimator fully utilizing longitudinal information is also consistent and asymptotically normally distributed. In contrast to $(\hat{a},\hat{b},\hat{b}^c,\hat{\lambda}_0,\hat{\lambda}_0^c)$, estimator $(\hat{a}^l,\hat{b}^l,\hat{b}^{c,l},\hat{\lambda}_0^l,\hat{\lambda}_0^{c,l})$ turns out to be more efficient. The details are summarized into the following theorem:

\begin{theorem}\label{theorem consistency full}
Under $\mathbf{C1}$-$\mathbf{C7}$,
the estimator $(\hat{a}^l,\hat{b}^l,\hat{b}^{c,l},\hat{\lambda}_0^l,\hat{\lambda}_0^{c,l})$ is consistent and asymptotically normally distributed, with their asymptotic variance being $1/m$ in scale of the asymptotic variance of $(\hat{a},\hat{b},\hat{b}^c,\hat{\lambda}_0,\hat{\lambda}_0^c)$, where $m$ is the number of censoring intervals.
\end{theorem}

Proof of theorem \ref{theorem consistency} and \ref{theorem consistency full} replies on three technical lemmas \ref{lemma1}-\ref{lemma3} that are stated in Appendix. 

\begin{remark} 
The proof of theorem \ref{theorem consistency}, \ref{theorem consistency full} and lemma \ref{lemma1}-\ref{lemma3} are quite technical, but the idea behind them are straightforward. Notice that the simulation-based estimator developed in this study is essentially a maximum likelihood estimator, so as long as the model \eqref{joint model} is identifiable and the standard regularity condition $\mathbf{C3}$-$\mathbf{C5}$ hold, consistency and asymptotic normality of our parametric estimator, $(\hat{a},\hat{b},\hat{b}^c)$, is just a consequence of the asymptotic property of the standard maximum likelihood estimator. As for the non-parametric part, $(\hat{\lambda}_0,\hat{\lambda}_0^c)$, its consistency still holds by the fact that when a model is identifiable, the true model setup is the unique maximal point of the entropy function which is a function in $\Omega$ \citep{amemiya1985advanced}. As for the asymptotic normality of non-parametric estimator, $(\hat{\lambda}_0,\hat{\lambda}_0^c)$, the proof is essentially the same as the proof in \cite{zheng2018understanding}. 

Therefore, model identifiability of \eqref{joint model} and the convergence of empirical likelihood function \eqref{likelihood} is the key to establish theorem \ref{theorem consistency} and \ref{theorem consistency full}, it is guaranteed by Lemma \ref{lemma1}-\ref{lemma3}. 

In fact, lemma \ref{lemma1}, combining with $\mathbf{C1}$ and $\mathbf{C2}$, provides a foundation to lemma \ref{lemma3} and model identifiability. This is done through verifying inequality \eqref{ineq} in an inductive way where condition $\mathbf{C2}$ is applied repeatedly to remove the possibility of existence of an invariant probability measure. 

Lemma \ref{lemma2} is crucial to the convergence of likelihood function. It helps confirm that samples generated from algorithm \ref{main algorithm 0}-\ref{main algorithm 3} are drawn correctly and satisfy the i.i.d. property, which guarantee that the empirical joint pdf \eqref{empirical pdf} approaches to the theoretical pdf \eqref{joint density 0 stage}, \eqref{joint density 1 stage} as the simulation sample size $N\rightarrow \infty$. In addition, with appropriate choice of tuning parameter $h,\,dt,\,N$ subject to condition $\mathbf{C7}$, lemma \ref{lemma2} also guarantees the convergence of likelihood function \eqref{likelihood} to the theoretical entropy function.

\end{remark}

\section{Numerical Studies}\label{numerical study}
\subsection{Simulation studies}
In this section, we present an example based on simulation studies to assess the finite sample performance of the proposed method. 
\begin{example}\label{example1}
200 random samples, each consisting of $n=100,\,200$ subjects, are generated from the model with $\mathfrak{p}=7$ dimensional longitudinal process with the $7$th dimension being the counting dimension. The six absolutely continuous dimensions are given as a special case of \eqref{bio-stat longi} where the error $e$ and random effect $\beta$ are supposed to be independent normal random vectors and follow $N(\mathcal{M}_1,\Sigma_1)$ and $N(\mathcal{M}_2,\Sigma_2)$, with the mean and co-variance matrix parametrized in the following way:
\begin{equation}\label{mu}
\mathcal{M}_i=(\mu_{i1},\dots,\mu_{i6})^{\top}
\end{equation}
\begin{equation}
\Sigma_i=\varA{diag}(\sigma^2_{i1},\dots,\sigma^2_{i6})
\end{equation}\label{sigma}
\noindent where $i=1,2$, $\varA{diag}(\cdot)$ denotes the diagonal matrix with diagonal elements given as the vector $\cdot$. In this example, we take $\mu_{ik}\equiv 0$ and $\sigma_{ik}\equiv 1$ for all $i$ and $k$, which means both the random effect and error are standard normal random vectors in this example. The configuration matrix $\mathcal{Z}_1\equiv0$ and $\mathcal{Z}_2(t)=t$. 

\noindent The counting dimension has its jumping hazard specified through \eqref{jump intensity} such that $b^c=(0,0,0,-1,1,0.6,1)$ and
\begin{equation}\label{lambda0c}
\lambda_0^c(t)=\frac{\exp(-3)+\exp^{-0.5t}}{\exp(-3)+1}.
\end{equation}
Finally, the hazard of terminal events is given by \eqref{joint model} where $b=(1,-1,0.3,0,0,0,1)$ and $\lambda_0$ satisfies that
\begin{equation}\label{lambda0}
\lambda_0(t)=\frac{\exp(-1)+\exp^{-t}}{\exp(-1)+1}.
\end{equation}

\noindent In sum, the model setup to be estimated for this example consist of three parts, the longitudinal parameters $(\mathcal{M}_1,\mathcal{M}_2,\Sigma_1,\Sigma_2)$, the parameter for counting dimension and terminal event $(b^c,b)$ and the non-parametric baseline hazards $(\lambda_0^c,\lambda_0)$. 
\end{example}

In estimation stage, for simplicity, we select $dt=1/n$, $N=n$ and $b=n^{-\frac{1}{2}}$, which are naturally compatible with the requirement of condition $\mathbf{C7}$. The combination of simulation algorithm \ref{main algorithm 0} and \ref{main algorithm 2} is utilized to generate simulatible sequence of longitudinal process, then algorithm \ref{main algorithm 3} is applied to generate random samples of $(Z_T,T)$ as well as the empirical likelihood \eqref{likelihood}.

The estimation performance are presented and evaluated through the bias, SSE and CP of longitudinal parameters, regression coefficients $b$, $b^c$ and the baseline hazard $\int_0^t\lambda_0(s)ds$, $\int_0^t\lambda_0^c(s)ds$, where bias is the sample mean of the estimate minus the true value, SSE is the sampling standard error of the estimate, and CP is the empirical coverage probability of the 95\% confidence interval. Results are collected in Table \ref{table: 1} for longitudinal parameters, in Table \ref{table: 2} for parameters of counting dimension and terminal event, and in Figure \ref{fig: fitting} for the non-parametric component.

By the fitting performance in Table \ref{table: 1}, \ref{table: 2} and Figure \ref{fig: fitting}, we conclude that in example \ref{example1} both of the fitting to parametric and non-parametric components are pretty good for both sample size. Meanwhile, there is no significant difference between $n=100$ and $n=200$, in terms of the estimation bias, variance and 95\% credential interval, which implies that the convergence predicted in theorem \ref{theorem consistency} arrives quite fast. So our estimation procedure can even work well for a relatively small sample ($n=100$).  

\subsection{Real data examples}\label{real data examples}
In this section, we apply our method to the consumer-loan data collected from one of the biggest P2P loan platform, renrendai.com, in China. This platform provides loan-level details regarding interest rate, loan amount, repayment term and the borrowers' credit records (the number of previous default, overdue, pre-payment, the number of applications, approved applications and pay-offs, credit score), the capacity of repayment (income, whether or not have unpaid mortgage, car loan, collateral situation) and the other miscellaneous background information of borrowers (education level, residential location, job position and industry, description of loan purpose). In addition, the repayment details are also included, such as the repayment amount, the number of anomalous repayment actions (pre-pay, overdue and default). To reflect the macro-economic condition during repayment, we also collect the province-level economic data such as GDP and housing price from National Bureau of Statistics of China.

We collect a sample of loans that were originated after Jan. 2015 from renrendai.com (this is because the publicly available  historical record of economic variables can only be traced back to Jan. 2015), and there are almost 225,000 of loan records. Terms of these loans vary from 1 year (12 months) to 3 years (36 months). About 1/5 of the loans have been terminated by the data collection time (2018 Jul.). Among the loans that have already been terminated, only a tiny portion are default or experienced over-due, the vast major portion are either paid-off or pre-paid before the declared term. In addition, among those pre-paid loans, more than two-third of them was closed within 8 months. Therefore, we select whether the loan has been pre-paid-off as the terminal event in interest, the final censoring time is the 8th-month (censor bound $C$ is set to $8$ in algorithm \ref{main algorithm 3}). Finally, after removing the records with missing attribute, there are 29,898 loans remained, which consist of the entire sample. In this real data example, we consider a $20$-dimensional covariate $Z$, which consists of 17 stationary variable (including loan-level and borrower-level variables) and 2 absolutely continuous processes that are monthly GDP and housing price, and 1 counting process, the number of anomalous repayments occurred during repayment period which is available only once, thus is the only covariate that has missing longitudinal observation.

Since longitudinal observations for macro-economic processes, such as GDP and housing prices, are always available in a constant frequency (e.g. monthly), the monthly observations have already formed a simulatible sequence to the underlying stochastic processes. Given a simulatible sequence for absolutely continuous components, the simulatible sequence to the full longitudinal processes can be easily generated by algorithm \ref{main algorithm 2}. Then the joint samples of $(W_S,S)$ and empirical likelihood function \eqref{likelihood} are built up from algorithm \ref{main algorithm 3}.

It turns out that our estimation procedure can be easily combined with variable selection techniques such as LASSO and adaptive LASSO \citep{tibshirani1996regression,zou2006adaptive}. Due to the substantial amount of covariates existing for renrendai data, we adopt an adaptive LASSO approach together with our simulation-based procedure to estimate model setup and identify the real significant variables involved in hazard function \eqref{joint model} and \eqref{jump intensity}. 

Dividing every observation time by censor bound $C=8$, we normalize the survival time scale to $1$ in the estimation and fitting plot. The fitting results are displayed in Table \ref{table: 3} and Figure \ref{fig: real}, where Table \ref{table: 3} records the estimated regression coefficients for both of the terminal event and counting covariate, where variable selection is done through adaptive LASSO. Figure \ref{fig: real} shows the point-wise fitting and empirical 95\% deviation (calculated in bootstrap way) of the cumulative baseline hazard of $\int_0^t\lambda_0(s)ds$ and $\int_0^t\lambda_0^c(s)ds$.

Table \ref{table: 3} shows that only a few portion of covariate variables have significantly impact on the prepayment behavior and the number of irregular payments (e.g. over-due) before pay-off, meanwhile the influential factors on these two events are distinct. For prepayment, both of housing price and GDP are quite influential, their influence are positive, indicates that when local economy goes up, borrowers tend to have good liquidity and prefer to pay off the loans in prior to expiration so as to reduce the total financial cost. The type of job position of borrowers can also influence their repayment behavior. For borrowers with relative low job positions, such as clerks, their preference to pre-paying off debt is strong, which implies that this group of borrowers are more sensitive to financial cost. In contrast, self-employed and manager-level borrowers are more likely to have irregular repayments, this observation might be relevant to their relatively unstable cash-flow.

\section{Discussion}\label{conclusion}
In this paper we proposed a simulation-based approach  to estimate the joint model of longitudinal and event-time data with a mixed longitudinal process with absolutely continuous components as well as a counting component. Our approach can generate well-performed estimators under the minimal availability condition of longitudinal observations, namely it allows missing longitudinal observations for part of or all of the temporal covariates before event time. So, our approach outperforms most of the existing semi-parametric estimation procedure in its flexibility to accommodate missing data.

In addition, the estimator generated from our approach is essentially an MLE-class estimator, but unlike the alternative method in literature, there is no latent variable involved in the likelihood function. So the computation of our procedure does not rely on the EM algorithm and can be embarrassingly parallelized, which effectively extends the applicability of our approach to massive financial data.

The limitation of our method is the ignorance of unbounded-variation longitudinal processes, such as Brownian motion and its variants, and the issue of how to fully utilize longitudinal information to lift estimation efficiency when longitudinal observation time is irregularly and sparsely distributed, which need to be handed in future studies.

\vspace{2cm}
\noindent{\large\bf Acknowledgements:} This work was partially supported by the Major and Key Technologies Program in Humanities and Social
Sciences of the Universities from Zhejiang Province (18GH037).

\begin{appendix}
\section{Proof of Theorem \ref{theorem consistency} \& \ref{theorem consistency full}}
Proof of Theorem \ref{theorem consistency} is complicated and need three lemmas stated as below. Proof for Theorem \ref{theorem consistency full} is essentially identical to that for Theorem \ref{theorem consistency}, therefore is omitted from the main body of the paper.  

\begin{lemma}\label{lemma1}
$ $ \\
\noindent (1). For every joint model \eqref{joint model} with the absolutely continuous components of its longitudinal process specified as \eqref{time-integral longi}, one counting component specified as \eqref{jump intensity}, the joint pdf of $(Z_T,T)$ is expressible through the following iterative process: 

\noindent At $j=0$,
\begin{equation} \label{joint density 0 stage}
p\left(j,z,t\mid \Omega\right)=
p_Z\left(j,z,t\mid a\right)\cdot
\rho_c\left(j,z,t\mid a,b^c,\lambda_0^c\right)\cdot \rho\left(j,z,t\mid a,b,\lambda_0\right)\cdot \exp(b^{-\ast\top}z)\lambda_0(t)
\end{equation}
\begin{equation} \label{jump density 0 stage}
p^c\left(j,z,t\mid \Omega\right)=p_Z\left(j,z,t\mid a\right)\cdot
\rho_c\left(j,z,t\mid a,b^c,\lambda_0^c\right)\cdot \exp((b^{c,-\ast})^{\top}z)\lambda_0^c(t)
\end{equation}
\noindent At $j>0$:
\begin{equation}\label{joint density 1 stage}
\Scale[0.8]{
\begin{aligned}
p\left(j,z,t\mid \Omega\right)= &\exp(b^{-\ast\top}z+b^{\ast} j)\lambda_0(t)\cdot
\int_{\mathbb{R}^{\mathfrak{p-1}}}
\int_0^{t}p^c\left(j-1,z',s\mid \Omega\right)\cdot
\rho_c\left(j,z',z,s,t\mid a,b^c,\lambda_0^c\right)\cdot \rho\left(j,z',z,s,t\mid a,b,\lambda_0\right)dsdz'\\
&+p_Z\left(j,z,t\mid a\right)\cdot
\rho_c\left(j,z,t\mid a,b^c,\lambda_0^c\right)\cdot \rho\left(j,z,t\mid a,b,\lambda_0\right)\cdot \exp(b^{-\ast\top}z+b^{\ast}j)\lambda_0(t)
\end{aligned}}
\end{equation}
\begin{equation}\label{jump density 1 stage}
\Scale[0.8]{
\begin{aligned}
p^c\left(j,z,t\mid \Omega\right)= &\exp((b^{c,-\ast})^{\top}z+b^{c,\ast} j)\lambda_0^c(t)\cdot\int_{\mathbb{R}^{\mathfrak{p-1}}}\int_0^{t}p^c\left(j-1,z',s\mid \Omega\right)\cdot
\rho_c\left(j,z',z,s,t\mid a,b^c,\lambda_0^c\right)dsdz'\\
&+p_Z\left(j,z,t\mid a\right)\cdot
\rho_c\left(j,z,t\mid a,b^c,\lambda_0^c\right)\cdot  \exp((b^{c,-\ast})^{\top}z+b^{c,\ast}j)\lambda_0^c(t)
\end{aligned}}
\end{equation}
where $\Omega=(a,b,b^c,\lambda_0,\lambda_0^c)$ is a given model setup, $p_Z(j,\cdot,t\mid a)$ is a pdf on $\mathbb{R}^{\mathfrak{p}-1}$ induced by absolutely continuous components of longitudinal process given $t$ and $j$, $p^c(j,z,t\mid \Omega)$ represents the joint density of the event that the counting component $Z^{\ast}$ jump from $j$ to $j+1$ at $t$ and $Z^{-\ast}(t)=z$. Denote supscript $\ast$, $-\ast$ as indicator to the counting component and absolutely continuous components of longitudinal process respectively, conditional probability function of $\rho$, $\rho_c$, $\rho'$ and $\rho_c'$ are defined as below:
\begin{align}\label{condition prob}
&\rho(j,z,t\mid a,b,\lambda_0)=\prob*{T>t| Z^{\ast}(0)=j,Z^{-\ast}(t)=z,a,b,\lambda^0}\\
&\rho_c(j,z,t\mid a,b^c,\lambda_0^c)=\prob*{Z^{\ast}(t)=j|\begin{aligned} 
&Z^{\ast}(0)=j,\\
&Z^{-\ast}(t)=z,\\
&a,b^c,\lambda_0^c
\end{aligned}}\\
&
\rho'(j,z',z,s,t\mid a,b,\lambda_0)=\prob*{T>t|
\begin{aligned} 
&Z^{\ast}(s)=
Z^{\ast}(t)=j,\\
&Z^{-\ast}(s)=z',\\
&Z^{-\ast}(t)=z,\\
&a,b,\lambda^0
\end{aligned}}\\
&
\rho_c'(j,z',z,s,t\mid a,b^c,\lambda_0^c)=\prob*{Z^{\ast}(t)=j|
\begin{aligned} 
&Z^{\ast}(s)=Z^{\ast}(t)=j,\\
&Z^{-\ast}(s)=z',\\
&Z^{-\ast}(t)=z,\\
&a,b^c,\lambda_0^c
\end{aligned}}
\end{align}

\noindent (2). In addition, under condition $\mathbf{C5}$, function $p_Z$ can be expressed as below:
\begin{equation}\label{p-tilde}
p_Z(j,z,t\mid \Omega)=p_Z\left(j,g\left(j,z,t,t\mid \Omega\right),0\mid\Omega\right)\cdot \mathcal{J}_{z|a}(t),
\end{equation}
The function $p_Z(j,z,0\mid \Omega)$ (treated as a function in variable $z$) is the initial pdf induced by $Z^{-\ast}_0$. For every $t$, $\mathcal{J}_{z|\Omega}(t)$ denotes the Jacobian of the function $g\left(j,z,t,t\mid a\right)$ (in the variable $z$) evaluated at the point $z$.   
The
function $g$ is determined by conditional expectation $q$ \eqref{function-q} through solving a family
of initial value problems (IVPs). Namely for every fixed $z$ and $t$,
$g\left(j,z,t,.\mid a\right)$ is the solution to the following ordinary differential
equation (ODE) for $s\in\left(0,t\right)$:
\begin{equation}\label{ode}
z'\left(s\right) = -q\left(j,z\left(s\right),t-s\mid a\right)
\end{equation}
subject to the initial condition $g\left(j,z,t,0\right)=z$.
 
\noindent (3). Functions  $\rho$, $\rho_c$, $\rho'$ and $\rho_c'$ satisfies that:
\begin{align}
&\Scale[0.8]{\begin{aligned}
&\rho(j,z,t\mid a,b,\lambda_0)\\
=&\expect*{\exp\left(-\int_0^t\exp(b^{-\ast\top}Z^{-\ast}(s)+b^{\ast}j)\lambda_0(s)ds\right)|
\begin{aligned} 
&Z^{-\ast}(t)=z,\\
&Z^{\ast}(t)=Z^{\ast}(0)=j,\\
&a,b,\lambda_0\end{aligned}}\end{aligned}}\\
&\Scale[0.8]{\begin{aligned}
&\rho_c(j,z,t\mid a,b^c,\lambda_0^c)\\
=&\expect*{\exp\left(-\int_0^t\exp((b^{c,-\ast})^{\top}Z^{-\ast}(s)+b^{c,\ast}j)\lambda_0^c(s)ds\right)|
\begin{aligned} &Z^{-\ast}(t)=z,\\
&Z^{\ast}(t)=Z^{\ast}(0)=j,\\
&a,b^c,\lambda_0^c\end{aligned}}\end{aligned}}\\
&\Scale[0.8]{\begin{aligned}
&\rho'(j,z',z,s,t\mid a,b^c,\lambda_0^c)\\
=&\expect*{\exp\left(-\int_s^t\exp(b^{-\ast\top}Z^{-\ast}(\tau)+b^{\ast}j)\lambda_0(\tau)d\tau\right)|
\begin{aligned}
&Z^{-\ast}(t)=z,\\
&Z^{-\ast}(s)=z',\\
&Z^{\ast}(t)=Z^{\ast}(s)=j,\\
&a,b,\lambda_0\end{aligned}}\end{aligned}}\\
&\Scale[0.8]{\begin{aligned}
&\rho_c'(j,z',z,s,t\mid a,b^c,\lambda_0^c)\\
=&\expect*{\exp\left(-\int_s^t\exp((b^{c,-\ast})^{\top}Z^{-\ast}(\tau)+b^{c,\ast}j)\lambda_0^c(\tau)d\tau\right)| \begin{aligned} &Z^{-\ast}(t)=z,\\
&Z^{-\ast}(s)=z',\\
&Z^{\ast}(t)=Z^{\ast}(s)=j,\\
&a,b^c,\lambda_0^c\end{aligned}}\end{aligned}}
\end{align}
\end{lemma}

\begin{lemma}\label{lemma2}
$ $ \\
(1). For an absolutely continuous longitudinal process with Markovian property \eqref{markovian}, $Z$, denote $\{(\zeta_{1,k},\dots,\zeta_{N,k}):\,k\in\mathbb{N}\}$ as the sequence generated from algorithm \eqref{main algorithm 1} with respect to fixed $dt$, $N$, the initial density $p_Z(\cdot,0)$ and the conditional density $p(\epsilon(t)\in x+dx\mid Z(t),t)$, a stochastic process on the discrete state space $\{1,\dots,N\}$ as below:
\begin{equation}\label{dis emp}
S(i,t)=s_{i,\zeta_{i,\left\lfloor t/dt\right\rfloor}}
\end{equation}
then the following condition holds as $dt\rightarrow 0$ and $N\rightarrow \infty$ for every finite integer $m$, every sequence sequence $t_1<\dots<t_m$ and every $m$-dimensional continuous function $f$:
\begin{equation}\label{weak conv}
\frac{1}{N}\sum_{i=1}^Nf(S(i,t_1),\dots,S(i,t_m))\rightarrow E(f(Z(t_1),\dots,Z(t_m)))
\end{equation}
In the other words, algorithm \ref{main algorithm 2} generate empirically simulatible sequences for every Markovian absolutely continuous longitudinal process $Z$.

\noindent (2). For a longitudinal process with one counting component and its absolutely continuous components satisfying Markovian property \eqref{markovian}, $Z=(Z^{\ast},Z^{-\ast})$, let $\zeta=\{(\zeta_{1,k},\dots,\zeta_{N,k}):\,k\in\mathbb{N}\}$ be the sequence generated from algorithm \eqref{main algorithm 1} with respect to fixed $dt$, $N$, the initial density $p_Z(\cdot,0)$ and the conditional density $p(\epsilon(t)\in x+dx\mid Z(t),t)$. The stochastic process constructed in \eqref{dis emp} with respect to $\zeta$ satisfy the weak convergence property \eqref{weak conv} as well, so algorithm \ref{main algorithm 3} generate empirically simulatible sequences for process $(Z^{\ast},Z^{-\ast})$.

\noindent (3). Algorithm \ref{main algorithm 3} generate i.i.d. samples of $(Z_T,T)$ for every fixed model setup $\Omega$.
\end{lemma}

\begin{lemma}\label{lemma3}
Under condition $\mathbf{C1}$ and $\mathbf{C2}$, for every model setup $\Omega\not=\Omega_0$, the joint pdf of $(Z_T,T)$ associated with $\Omega$ is not identical to that associated with $\Omega_0$, i.e.
\begin{equation}\label{ineq}
p(j,z,t\mid \Omega_0)\not=p(j,z,t\mid \Omega)
\end{equation} for some $(j,z,t)$ in their domain. In the other words, joint model \eqref{joint model} is identifiable.
\end{lemma}

\subsection{Proof of Lemma \ref{lemma1}}

Throughout this proof, we will suppress $\Omega$ from the arguments of function $\rho$, $p_Z$, $p^c$, $p$ and $q$ because they are all fixed constant. In addition, when necessary, we will use supscript $^{\ast}$ and $^{-\ast}$ to represent the value associated with counting component and absolutely continuous components, respectively. 

The proof is decomposed to two parts, in the first part, we verify the statement (2) and (3) in lemma \ref{lemma1}. In the second part, we decompose the proof of statement (1) to two cases and firstly validate the expression \eqref{joint density 0 stage}-\eqref{jump density 1 stage} in a simpler case where no counting component are involved in the longitudinal process. Then, we extend the proof of the simple case to the complete case with counting component added.

\begin{proof}[{\bf Part 1}:]
For statement {\bf (2)}, notice that
for every $j\in\mathbb{N}$ and $f\in C_{0}^{1}\left(\mathbb{R}^{\mathfrak{p}-1}\right)$, the following holds:\\
\begin{equation*}
\Scale[0.8]{
\begin{aligned}
\textrm{E}\left(f\left(Z^{-\ast}(t)\right)\right) = & \textrm{E}\left(f\left(Z^{-\ast}_0\right)\right)+
\textrm{E}\left(\intop_{0}^{t}f'\left(Z^{-\ast}(s)\right)
\epsilon_{s}ds\right)\\
 = & \textrm{E}\left(f\left(Z^{-\ast}_{0}\right)\right)+\intop_{0}^{t}
 \textrm{E}\left(f'\left(Z^{-\ast}(s)\right)\epsilon(s)\right)ds\\
 = & \textrm{E}\left(f\left(Z^{-\ast}_{0}\right)\right)+
\intop_{0}^{t}\textrm{E}\left(f'\left(Z^{-\ast}(s)\right)
q\left(j,Z^{-\ast}(s),s\right)\right)ds\\
 = & \textrm{E}_{Z^{-\ast}_{0}}\left(f\left(z\right)\right)+
 \intop_{0}^{t}\textrm{E}_{Z^{-\ast}(s)}\left(f\left(z\right)
 q\left(j,z,s\right)\right)ds
 \end{aligned}}
\end{equation*}
If we define the following operator $\mathcal{A}$ over the set of all continuous-time
processes:

\[
\mathcal{A}Z^{-\ast}(t):=Z^{-\ast}_{0}+\intop_{0}^{t}q\left(j,Z^{\ast}(s),s\right)ds
\]
then, the process $\left\{ \mathcal{A}Z^{\ast}(s)\right\} $ satisfies for all $f\in C_{0}^{1}\left(\mathbb{R}^{\mathfrak{p}-1}\right)$:\\
\begin{equation*}
\Scale[0.8]{
\begin{aligned}
\textrm{E}\left(f\left(\mathcal{A}Z^{-\ast}(t)\right)\right) = & \textrm{E}\left(f\left(Z^{-\ast}_{0}\right)\right)+
\intop_{0}^{t}
\textrm{E}\left(f'\left(Z^{-\ast}(s)\right)
q\left(j,Z^{-\ast}(s),s\right)\right)ds\\
  = & \textrm{E}_{Z^{-\ast}_{0}}\left(f\left(z\right)\right)+
\intop_{0}^{t}\textrm{E}_{Z^{-\ast}(s)}\left(f'\left(z\right)
q\left(j,z,s\right)\right)ds\\
  = & \textrm{E}\left(f\left(Z^{-\ast}(t)\right)\right)\end{aligned}}
\end{equation*}
Therefore, we have the following fact,

{\bf Fact}: For every fixed $j$, if $\{Z^{-\ast}(t)\}$ is an absolutely continuous process, $\{\mathcal{A}Z^{-\ast}(t)\}$
is a process equivalent to $\{Z(t)\}$ in terms of the collection of
induced pdf $p_{Z}(j,\cdot,t)$ for all $t$.
Consequently,
\[
\begin{aligned}
\textrm{E}_{Z^{-\ast}_{0}}\left(f\left(z\right)\right)&+
\intop_{0}^{t}\textrm{E}_{\mathcal{A}Z^{-\ast}(s)}\left(f'\left(z\right)
q\left(j,z,s\right)\right)ds=\\
&\textrm{E}_{Z^{-\ast}_{0}}
\left(f\left(z\right)\right)+\intop_{0}^{t}
\textrm{E}_{Z^{-\ast}(s)}\left(f'\left(z\right)q\left(j,z,s\right)\right)
ds\end{aligned}.
\]
By induction, for every $f\in C_{0}^{1}\left(\mathbb{R}^{\mathfrak{p}-1}\right)$:
\begin{equation}\label{iteration}
\Scale[0.8]{
\begin{aligned}
&\textrm{E}\left(f\left(\mathcal{A}^{n+1}Z^{-\ast}(t)\right)\right)  \\
=&  \textrm{E}\left(f\left(Z^{-\ast}_{0}\right)\right)+
\intop_{0}^{t}\textrm{E}\left(f'\left(\mathcal{A}^{n}Z^{-\ast}(s)\right)
q\left(j,\mathcal{A}^{n}Z^{-\ast}(s),s\right)\right)ds\\
  =&  \textrm{E}_{Z^{-\ast}_{0}}\left(f\left(z\right)\right)+
\intop_{0}^{t}\textrm{E}_{\mathcal{A}^{n}Z^{-\ast}(s)}
\left(f'\left(z\right)q\left(j,z,s\right)\right)ds\\
  =&  \textrm{E}_{Z^{-\ast}_{0}}\left(f\left(z\right)\right)+
\intop_{0}^{t}\textrm{E}_{Z^{-\ast}(s)}\left(f'\left(z\right)
q\left(j,z,s\right)\right)ds\\
  = & \textrm{E}\left(f\left(Z^{-\ast}(t)\right)\right)
 \end{aligned}}
\end{equation}
Consequently, if we start from a process $\{Z^{-\ast}(t)\}$,
by iteration of the operator $\mathcal{A}$, we would always get a process equivalent
to $\{Z^{-\ast}(t)\}$ in its 1-dimensional marginal pdf for every $t$. By the existence theorem of the solution
to an initial value problem associated with $q$ (\cite{perko2013differential}), we know that the sequence of processes $\left\{ \mathcal{A}^{n}Z^{-\ast}(t)\right\} _{n=0}^{\infty}$
converges point-wisely to some degenerated process satisfying 
\begin{equation}\label{YD}
Z^{\mathcal{D}}(t)=Z_{0}+\int_{0}^{t}q\left(j,
Z^{\mathcal{D}}(s),s\right)ds.
\end{equation} The point-wise convergence implies the
equivalence in distribution $\{ p_Z(j,\cdot,t):\, t\in[0,\infty)\} $
between the limit process and the initial process $Z^{-\ast}(t)$.

Obviously the above integral equation is equivalent to the time reversal of the initial value problems \eqref{ode}.
Solving that equation and applying the change of variable formula to its solutions
curves $g$, it is verified that $p_Z(j,\cdot,t)$ has the expression \eqref{p-tilde}. This completes the proof for statement (2).

For statement {\bf (3)}, using the definition of the Cox hazard function \eqref{joint model} in \cite{cox1972regression}, for every fixed $j$, $s<t$ and trajectory $Z_{\omega}(t)$ of longitudinal process $Z$ such that $Z_{\omega}(t)=z$, $Z_{\omega}(s)=z'$ and $Z^{\ast}_{\omega}(\tau)\equiv j$, the following relation holds by \cite{andersen1982cox,andersen1992repeated}:
\begin{equation}\label{condition on single traj}
\textrm{Pr}\left(T>t\mid Z_{\omega}(\tau),\, \tau \in [s,t) \right)=\exp\left(-\int_s^t \lambda(\tau,Z_{\omega}(\tau))d\tau\right)
\end{equation}
Through taking conditional expectation for both sides of \eqref{condition on single traj} with respect to all longitudinal trajectories satisfying $Z^{\ast}(\tau)\equiv j$, $Z^{-\ast}(s)=z'$ and $Z^{-\ast}(t)=z$, the relation for $\rho'$ in statement (3) is established. By similar argument, the remaining three relations in statement (3) can be verified. 
\end{proof}

\begin{proof}[{\bf Part 2}:]
In the second part of the proof, we verify the statement (1) of lemma \ref{lemma1}. In this part, we consider two cases:\\
\noindent i) the longitudinal process only consists of absolutely continuous components;\\
\noindent ii) there exist one extra counting component.

For case {\bf i)}, firstly, notice that it is equivalent between that the longitudinal process has $\mathfrak{p}$ dimension and involve no counting component and that the longitudinal process has $\mathfrak{p}+1$-dimensional with one counting component while the counting component is constantly zero. So, lemma \ref{lemma1} holds for case (1) if and only if the joint pdf of $(Z_T,T)$ in case (1) has exactly the form \eqref{joint density 0 stage} with $\rho_c\equiv 1$. To verify this, we consider probability of occurrence of the following event:
\begin{equation}\label{eventA}
A_{z,t,\delta}:=\textrm{Pr}\left\{ Z(t)\in\left(z-\delta,z\right),T<t\right\} 
\end{equation}
By definition of function $\rho$ in \eqref{condition prob}, it is obvious that the viability of joint pdf \eqref{joint density 0 stage} is equivalent to the establishment of the following identity:

\begin{equation} \label{A3_2}
A_{z,t,\delta}=\int_{z-\delta}^{z}\int_{0}^{t}\rho(z,s)\cdot \exp(b^{\top}x)\cdot p_Z\left(x,s\right)dsdx.
\end{equation}

To verify \eqref{A3_2}, notice that 
\begin{equation}
\Scale[0.8]{
\begin{aligned}
\left\{ Z(t)\in\left(z-\delta,z\right),T< t\right\}
= & \bigcap_{\Delta>0}\bigcup_{s_i\in S_n}\left\{ Z(s_{i})\in\left(z-\delta,z\right),s_{i}\leq T<s_{i+1}\right\} \\
= & \bigcap_{\Delta>0}\bigcup_{s_i\in S_n }\left(\left\{ Z(s_{i})\in\left(z-\delta,z\right),s_{i}\leq T\right\} /\left\{ Z(s_{i})\in\left(z-\delta,z\right),s_{i+1}\leq T\right\} \right),
\end{aligned}}
\end{equation}
where $S_n=\left\{ s_{i}=i\cdot\Delta:i=0,\dots,n,n\cdot\Delta\leq t<\left(n+1\right)\cdot\Delta\right\}$. Therefore,
\begin{equation*} \label{A3_5}
\begin{aligned}
A_{z,t,\delta} = &\Scale[0.8]{\lim_{\Delta\rightarrow0}\sum_{i=0}^{n_{t,\Delta}}\left(\textrm{E}\left(\mathbf{1}_{\left\{ Z(s_{i})\in\left(z-\delta,z\right)\right\} }\cdot\mathbf{1}_{\left\{ s_{i}\leq T\right\} }\right)-\textrm{E}\left(\mathbf{1}_{\left\{ Z(s_{i})\in\left(z-\delta,z\right)\right\} }\cdot\mathbf{1}_{\left\{ s_{i}+\Delta\leq T\right\} }\right)\right)}\\
 =&\Scale[0.8]{ \lim_{\Delta\rightarrow0}\sum_{i=0}^{n_{t,\Delta}}\left(\begin{aligned}
 &\textrm{E}\left(\mathbf{1}_{\left\{ Z(s_{i})\in\left(z-\delta,z\right)\right\} }\cdot \textrm{E}\left(s_{i}\leq T|Z(s_{i})\right)\right)-\\
 &\textrm{E}\left(\mathbf{1}_{\left\{ Z(s_{i})\in\left(z-\delta,z\right)\right\} }\cdot \textrm{E}\left(s_{i}+\Delta\leq T|Z(s_{i})\right)\right)\end{aligned}\right)}
\end{aligned}
\end{equation*}
Then, by the definition of function $\rho$ and $\rho'$ in \eqref{condition prob}, we have the identity \eqref{A3_conclusion} hold

\begin{equation}
\label{A3_conclusion}
\Scale[0.7]{
\begin{aligned}
&A_{z,t,\delta}\\  
= & \lim_{\Delta\rightarrow0}\sum_{i=0}^{n_{t,\Delta}}\left(
\begin{gathered}
\int_{z-\delta}^{z}\rho
\left(x,s_{i}\right)\cdot p_Z
\left(x,s_{i}\right)dx\\
-\\
\int_{z-\delta}^{z}\int_{0}^{\infty}\rho'
\left(x,x+\int_{s_{i}}^{s_{i}+\Delta}
\epsilon(\tau)
d\tau,s_{i},s_{i}+\Delta\right)
d\textrm{Pr}\left(\int_{s_{i}}^{s_{i}+
\Delta}\epsilon(\tau)d\tau|Z(s_{i})=x\right)\cdot\rho
\left(x,s_{i}\right)\cdot p_Z
\left(x,s_{i}\right)dx
\end{gathered}
\right)\\
= & -\lim_{\Delta\rightarrow0}\sum_{i=0}^{n_{t,\Delta}}\frac{\int_{z-\delta}^{z}\left(
\int_{0}^{\infty}\rho'\left(x,x+\int_{s_{i}}^{s_{i}+
\Delta}\epsilon(\tau)d\tau,s_{i},s_{i}+\Delta\right)-
1\right)d\textrm{Pr}\left(\int_{s_{i}}^{s_{i}+\Delta}\epsilon_{\tau}d\tau|Z(s_{i})=x\right) \cdot\rho
\left(x,s_{i}\right)\cdot p_Z\left(x,s_{i}\right)dx}{\Delta}\cdot\Delta\\
= & -\lim_{\Delta\rightarrow0}\sum_{i=0}^{n_{t,\Delta}}\frac{\int_{z-\delta}^{z}\int_{0}^{\infty}\left(
\partial_2 \rho'\cdot\int_{s}^{s+\Delta}\epsilon(\tau)d\tau+\partial_4 \rho'\cdot\Delta\right)\left(x,x,s_i,s_i\right)
d\textrm{Pr}\left(\int_{s_{i}}^{s_{i}+\Delta}\epsilon(\tau)d\tau|Z(s_{i})=x\right)\cdot\rho
\left(x,s_{i}\right)\cdot p_Z\left(x,s_{i}\right)dx}{\Delta}\cdot\Delta\\
= & -\lim_{\Delta\rightarrow0}\sum_{i=0}^{n_{t,\Delta}}\frac{\int_{z-\delta}^{z}\left(\partial_2
\rho'\left(x,x,s_i,s_{i}\right)\cdot
\int_{0}^{\infty}\left(\int_{s_{i}}^{s_{i}+
\Delta}\epsilon(\tau)d\tau\right)d\textrm{Pr}\left(\int_{s_{i}}^
{s_{i}+\Delta}\epsilon(\tau)d\tau|Z(s_{i})=x\right)+\partial_4\rho'\left(x,x,s_{i},s_{i}\right)\cdot\Delta\right) p_Z\left(x,s_{i}\right)dx}{\Delta}\cdot\Delta\\
= & -\lim_{\Delta\rightarrow0}\sum_{i=0}^{n_{t,\Delta}}\int_{z-\delta}^{z}\left(\partial_2\rho'
\left(x,x,s_i,s_{i}\right)\cdot \textrm{E}\left(\frac{\int_{s_{i}}^{s_{i}+\Delta}\epsilon(\tau)d\tau}{\Delta}|Z(s_{i})=x\right)+\partial_4\rho'\left(x,x,s_i,s_{i}
\right)\right)\cdot\rho
\left(x,s_{i}\right)\cdot p_Z\left(x,s_{i}\right)dx\cdot\Delta\\
= & -\lim_{\Delta\rightarrow0}\sum_{i=0}^{n_{t,\Delta}}\int_{z-\delta}^{z}\left(\partial_2\rho'
\left(x,x,x_i,s_{i}\right)\cdot \textrm{E}\left(\epsilon(s_{i})\mid Z(s_{i})=x\right)+\partial_4\rho'\left(x,x,s_is_{i}
\right)\right)\cdot\rho
\left(x,s_{i}\right)\cdot p_Z\left(x,s_{i}\right)dx\cdot\Delta\\
= & -\lim_{\Delta\rightarrow0}\sum_{i=0}^{n_{t,\Delta}}\int_{z-\delta}^{z}\left(\partial_2\rho'
\left(x,x,s_i,s_{i}\right)\cdot q\left(x,s_{i}\right)+\partial_2\rho'\left(x,x,s_i,
s_{i}\right)\right)\cdot\rho
\left(x,s_{i}\right)\cdot p_Z\left(x,s_{i}\right)dx\cdot\Delta\\
 = & \int_{z-\delta}^{z}\int_{0}^{t}-\left(\partial_2\rho'
 \left(x,x,s,s\right)\cdot q\left(x,s\right)+\partial_4\rho'\left(x,x,s,s\right)\right) \cdot\rho
\left(x,s_{i}\right)\cdot p_Z\left(x,s\right)dsdx,
\end{aligned}}
\end{equation}

\noindent where we use the relation $\rho'(x,x,s,s)\equiv 1$ by definition \eqref{condition prob}, $\partial_l$ refers to the partial derivative operator associated with the $l$th argument variable, $q$ is the conditional expectation function for $\epsilon(t)$ defined in \eqref{function-q} with its dependence on $j$ and $\Omega$ suppressed.
 
\eqref{A3_conclusion} implies
\begin{equation}\label{p}
\Scale[0.8]{
p(z,t)=-\left(\partial_2\rho'
 \left(x,x,s,s\right)\cdot q\left(x,s\right)+\partial_4\rho'\left(x,x,s,s\right)\right) \cdot\rho
\left(x,s_{i}\right)\cdot p_Z\left(x,s\right),
}
\end{equation}
from the definition of $\rho$ in \eqref{condition prob}, the meaning of Cox hazard function \eqref{joint model} in \cite{cox1972regression} and \eqref{p}, we have
\begin{equation}\label{substitute}
\Scale[0.8]{
\lambda_0(t,z)=\frac{p_Z\left(z,t\right)\cdot\rho(z,t)}{p(z,t)}=-\partial_2\rho'
 \left(x,x,s,s\right)\cdot q\left(x,s\right)-\partial_4\rho'\left(x,x,s,s\right)
}
\end{equation}
Therefore, combining the Cox hazard function in \eqref{joint model} and \eqref{substitute},
\begin{equation}
p(z,t)=p_Z(z,t)\cdot \rho(z,t)\cdot \exp(b^{\top}z)\lambda_0(t)
\end{equation}
that completes the proof for case {\bf i)}.

For case {\bf ii)}, it is derivable straightforwardly by an induction. First, when $j=0$, the joint pdf of \eqref{joint density 0 stage} is justifiable through the argument of case i) and the definition of $\rho_c$ in \eqref{condition prob}. In addition, regardless whether or not the terminal event occurs, the occurrence of event that the counting component increases by $1$ can be completely modelled by case i), so the argument in case i) is directly applicable to verify the expression \eqref{jump density 0 stage}, therefore, we complete the verification in the step $j=0$.

When $j>0$, for every fixed $z'$ and $s$, the integrand involved in the first term of \eqref{joint density 0 stage} is nothing more than the joint density of the following five events:\\
\indent 1. $Z^{-\ast}(t)=z$,\\
\indent 2. $T=t$,\\
\indent 3. $Z^{\ast}(t)=j$,\\
\indent 4. $Z^{-\ast}(s)=z'$,\\
\indent 5. $J_j=s$;\\
\noindent where $J_j$ is the jump time of counting component from stage $j-1$ into $j$. So, through integrating out $z'$ and $s$, the first term of \eqref{joint density 1 stage} gives the joint pdf of $(Z^{-\ast}_T=z,Z^{\ast}_T=j,T=t)$ when the system entered into the stage $Z^{\ast}=j$ at some time before $t$. In contrast, the second term of \eqref{joint density 1 stage} gives  the joint pdf of $(Z^{-\ast}_T=z,Z^{\ast}_T=j,T=t)$ when the system was initialized at the stage $Z^{\ast}=j$. Thus, adding the two terms together integrates out the effect of initialization, and returns the complete joint pdf of $(Z^{-\ast}_T=z,Z^{\ast}_T=j,T=t)$, so expression \eqref{joint density 1 stage} is validated. Analogously, the validity of \eqref{jump density 1 stage} can be demonstrated in exactly same way as long as we redefine the event in interest as transition of the counting component from $j$ to $j+1$. Proof for lemma \ref{lemma1} completes.

\end{proof}

\subsection{Proof of Lemma \ref{lemma2}}
\begin{proof}
In this proof, we firstly show statement (2) and (3) on the basis that statement (1) holds, then sketch the proof for statement (1).

Given statement (1) and the functional form \eqref{jump intensity} of the jump hazard, we can think of a simple version of the longitudinal process where the counting component can only jump once, i.e. the range of the counting component is a binary set $\{0,1\}$. In this case, statement (2) is verified directly by the result of statement (1) and the relation between hazard function and survival function \citep{andersen1982cox,andersen1992repeated}.

When multiple jumps exist, statement (2) can be verified by induction. In fact, when $j>1$, the longitudinal process with its counting component having at most $j$ jumps, denoted as $L_j$, is equivalent in distribution to the composition of a longitudinal process with its counting component having at most $j-1$ jumps, denoted as $L_{j-1}$, and a binary event process with its hazard function specified through \eqref{jump intensity}, while for $L_{j-1}$ an empirically simulatible sequence has been generated such that \eqref{weak conv} holds. Then, to verify \eqref{weak conv} for $L_j$, it suffices to show that the simulated occurrence of the binary event by algorithm \ref{main algorithm 2} at every time $t$ when $L_{j-1}(t)=z$ and $L_{j-1}(s)=z'$ asymptotically follows the correct joint probability, which, by \eqref{jump intensity} and the relation between hazard function and survival function \cite{andersen1982cox,andersen1992repeated}, must be represented as an integral of \eqref{condition on single traj} with respect to all trajectories end up with $z$ at $t$ and $z'$ at $s$. In the other word, it suffices to verify the identity \eqref{inductive verify}

\begin{equation}\label{inductive verify}
\Scale[0.8]{
\begin{aligned}
\frac{1}{N}\sum_{i=1}^N\exp&\left(-I(L_{i,j-1}(s+l\cdot dt)\in z+dz, L_{i,j-1}(s)\in z'+dz)\cdot\sum_{k=1}^l\exp
\left(b^{c\top}\cdot L_{i,j-1}(s+k\cdot dt)\right)\cdot \lambda_0^c(s+k\cdot dt)\right)\\
\rightarrow &
\textrm{E}\left(I( L_{\omega,j-1}(t)\in z+dz,L_{\omega,j-1}(s)\in z'+dz)\cdot \exp\left(-\int_s^t \lambda_0^c(\tau)\exp(b^{c\top}\cdot L_{\omega,j-1}(\tau))d\tau\right) \right)
\end{aligned}}
\end{equation}

as $dt\rightarrow 0$, $N\rightarrow\infty$, where $l$ is taken as the least integer such that $l\cdot dt+s>t$, $L_{\omega,j-1}$ denote a sample trajectory of longitudinal process $L_{j-1}$, $L_{i,j-1}$ denote the $i$th sample sequence generated from algorithm \ref{main algorithm 2} for $L_{j-1}$, $I$ denotes the indicator function. By induction assumption, \eqref{inductive verify} holds. For the case $j=1$, \eqref{inductive verify} has already been verified in the previous paragraph, so proof for statement (2) completes.

Note that the same argument in the proof for statement (2) is directly applied to prove statement (3).

\begin{remark}
The proof of statement (2) and (3) only relies on the conclusion of statement (1), but does not depend the Markovian condition \eqref{markovian} required by algorithm \ref{main algorithm 1}. In the other word, as long as an empirically simulatible sequence can be generated for absolutely continuous components of the longitudinal process, algorithm \ref{main algorithm 2} and \ref{main algorithm 3} are still applicable to generate the desired i.i.d. samples. So, in principle, the estimation proposed in this paper is extendible to more general settings.
\end{remark}

Finally, for statement (1), the weak convergence condition \eqref{weak conv} can be established in exactly the same way as the construction of numerical solutions to a stochastic differential equation, we refer the audience to the textbook \cite{karatzas2012brownian} for more details.
\end{proof}
\subsection{Proof of Lemma \ref{lemma3}}

\begin{proof}

The proof follows the induction steps of the construction of joint pdf in lemma \ref{lemma1}. For simplicity, we firstly consider the simple case that the longitudinal process assigns positive mass to but does not fully concentrate on the event $Z^{\ast}(0)=0$. Under this assumption, we will show that if exist some model setup $\Omega=(\tilde{a},\tilde{b},\tilde{b}^c,\tilde{\lambda}_0,\tilde{\lambda}_0^c)$ that also associates to the true joint pdf \eqref{joint density 0 stage}, then i) $\tilde{b}=b$, ii) $\tilde{\lambda}_0=\lambda_0$, iii) $\tilde{a}=a$, iv) $\tilde{b}^c=b^c$, v) $\tilde{\lambda}_0^c=\lambda_0^c$, where $a,\,b,\,b^c,\,\lambda_0,\lambda_0^c$ are the true model setup included in $\Omega_0$.\\

\noindent Proof for {\bf i)}, suppose $\tilde{b}\not=b$. Then, by the condition $\mathbf{C2}$ and the statement (1), (3) of lemma \ref{lemma1}, we have 
\begin{equation}\label{initial identity}
\Scale[0.8]{
p(j,z,0\mid \Omega_0)\exp(b^{-\ast\top}z+b^{\ast\top}j) \equiv p(j,z,0\mid \Omega)\exp(\tilde{b}^{-\ast\top}z+\tilde{b}^{\ast\top}j)
}
\end{equation}
Obviously, \eqref{initial identity} contradict to condition $\mathbf{C1}$ as long as $b\not=\tilde{b}$, which enforces that for all $\Omega\not=\Omega_0$ their projection to sub-coordinate $b$ must agree with $\Omega_0$. Notice that \eqref{initial identity} also implies when $\Omega\not=\Omega_0$ can induce the same joint pdf, the initial pdf of longitudinal processes must satisfy $p(j,z,0\mid \Omega_0)\equiv p(j,z,0\mid \Omega)$.\\

\noindent Proof for {\bf ii)}: suppose $\tilde{\lambda}_0(t)\not=\lambda_0(t)$, by i), \eqref{joint density 0 stage} and the assumption that $Z^{\ast}(0)=j$ is assigned with positive mass, we have the identity \eqref{condition-for-lambda} hold.

\begin{equation}\label{condition-for-lambda}
p_Z\left(0,z,t\mid \Omega_0\right)\cdot
\rho_c\left(0,z,t\mid \Omega_0\right)\cdot \rho\left(0,z,t\mid \Omega_0\right)\cdot \lambda_0(t)=p_Z\left(j,z,t\mid \Omega\right)\cdot
\rho_c\left(0,z,t\mid \Omega\right)\cdot \rho\left(0,z,t\mid \Omega\right)\cdot \tilde{\lambda}_0(t)
\end{equation}

Since $\Omega$ and $\Omega_0$ associate with the same joint pdf of $Z_T$ and $T$, identity \eqref{joint survival} must hold as well.

\begin{equation}\label{joint survival}
\Scale[0.8]{
\int_{\mathbb{R}^{\mathfrak{p}-1}}p_Z\left(0,z,t\mid \Omega_0\right)\cdot
\rho_c\left(0,z,t\mid \Omega_0\right)\cdot \rho\left(0,z,t\mid \Omega_0\right)dz\equiv  \int_{\mathbb{R}^{\mathfrak{p}-1}}p_Z\left(0,z,t\mid \Omega\right)\cdot
\rho_c\left(0,z,t\mid \Omega\right)\cdot \rho\left(0,z,t\mid \Omega\right)dz},
\end{equation}

This is because both sides of \eqref{joint survival} gives the survival function of $T>0$ when $Z^{\ast}\equiv 0$ which is completely determined by the joint pdf of $(Z_T,T)$ at the stage $0$. Hence, under the assumption that $\Omega$ and $\Omega_0$ corresponds to exactly the same joint pdf, the equation \eqref{condition-for-lambda} enforces that $\tilde{\lambda}_0=\lambda_0$.\\

\noindent Proof for {\bf iii)}: (3) Suppose existing $\tilde{a}\not=a$ for which the joint pdf \eqref{joint density 0 stage} is identical. Without loss of generality, we assume that when $t=0$, $\mathbf{C2}$ (ii) holds. In general, for $\mathbf{C2}$ (ii) holds at some $t^{\ast}>0$, the following proof is essentially the same, the only modification is to change replace $p_Z(0,z,0\mid \Omega_0)$ with $p_Z(0,z,t^{\ast}\mid \Omega_0)\cdot \rho(0,z,t\mid a,b,\lambda_0)\cdot\rho_c(0,z,t\mid a,b^c,\lambda_0^c)$. 

\noindent In fact, when joint pdf are identical for $a\not=\tilde{a}$, the \eqref{weak-identity} holds

\begin{equation}\label{weak-identity}
\Scale[0.9]{
\begin{aligned}
&p_Z\left(0,z,0\mid \Omega_0\right)\mathcal{J}_{g^{-1}(0,z,0,t|a)|a}(t)\rho(0,g^{-1}(0,z,0,t|a),t\mid\Omega_0)\rho_c(0,g^{-1}(0,z,0,t|a),t\mid\Omega_0)
\lambda_0(t)\\
&=
p_Z\left(0,\tilde{z},0\mid \Omega_0\right)\mathcal{J}_{g^{-1}\left(0,\tilde{z},0,t\mid \tilde{a}\right)|\tilde{a}}(t)\rho(0,g^{-1}\left(0,\tilde{z},0,t\mid \tilde{a}\right),t\mid\Omega)\rho_c(0,g^{-1}\left(0,\tilde{z},0,t\mid \tilde{a}\right),t\mid\Omega)
\tilde{\lambda}_0(t)).
\end{aligned}}
\end{equation}

for all pairs $(z,\tilde{z})$ such that $z=g\left(0,g^{-1}\left(0,\tilde{z},0,t\mid \tilde{a}\right),t,t\mid a\right)$ where $g^{-1}\left(0,z,s,t\mid a\right)$ is
the inverse trajectories of $g$ and defined through the relation
\begin{equation}\label{g-inverse}
g\left(g^{-1}\left(0,z,s,t\mid a\right),s+t,t\mid a\right)=z.
\end{equation} Factor out \eqref{weak-identity} by $\mathcal{J}_{g^{-1}(0,z,0,t|a)|a}(t)\rho(0,g^{-1}(0,z,0,t|a),t\mid\Omega_0)\rho_c(0,g^{-1}(0,z,0,t|a),t\mid\Omega_0)
\lambda_0(t)$ and take the limit as $t\rightarrow 0$ yielding the following identity:
\begin{equation}\label{invariant-measure}
p_Z(0,\mathcal{T}_{r}(z),0\mid \Omega_0)\cdot\mathcal{J}_{\mathcal{T}_r}(z)=p_Z(0,z,0\mid \Omega_0)
\end{equation} 
where for every $r\in \mathbb{R}$, the map $\mathcal{T}_{r}:\mathbb{R}^{p}\rightarrow \mathbb{R}^{p}$ is the diffeomorphism obtained from solving the ODE system:
\begin{equation}\label{ode2}
z'=q\left(0,z,0\mid a\right)-q\left(0,z,0\mid \tilde{z}\right),
\end{equation}
$\mathcal{T}_r(z_0)$ is just the point reached at the time $r$ by the trajectory starting at $z_0$ that solves \eqref{ode2}. $\mathcal{J}_{\mathcal{T}_r}$ is the Jacobian associated with $\mathcal{T}_r$. By the language of ergodic theory, Eq. \eqref{invariant-measure} implies that the probability measure corresponding to the initial pdf $p_Z(0,\cdot,0\mid \Omega_0)$ is invariant under the $\mathbb{R}$-action on the space $\mathbb{R}^{\mathfrak{p}-1}$ induced by $\mathcal{T}$, the family of solutions to \eqref{ode2}. However, under the condition $\mathbf{C2}$ (ii), the action $\mathcal{T}$ associated with the pair of $a$ and $\tilde{a}$ does not allow any invariant probability measure fully supported on $\mathbb{R}^{\mathfrak{p}-1}$ unless $\tilde{a}=a$. This contradiction guarantees the condition $\tilde{a}=a$.\\

\noindent Combining i) and iii), we have 
\begin{equation}\label{pz identity}
p_Z(j,z,t\mid\Omega)\equiv p_Z(j,z,t\mid \Omega_0)
\end{equation}
and the identity
\begin{align}\label{rho identity}
&\rho(\cdot\mid\Omega)\equiv\rho(\cdot\mid\Omega_0),\\ &\rho'(\cdot\mid\Omega)\equiv\rho'(\cdot\mid\Omega_0)
\end{align}
whenever $\Omega$ and $\Omega_0$ induce the same joint pdf of \eqref{joint density 0 stage} and \eqref{joint density 1 stage}. The identity for $\rho$ and $\rho'$ is because they are completely determined by $b$, $\lambda_0$, the trajectory information encoded in parameter $a$ and the prescribed stage where the counting component is on by the statement (3) of lemma \ref{lemma1}.\\

\noindent Proof for {\bf iv)} and {\bf v)}: Using identity \eqref{pz identity} and \eqref{rho identity}, and the assumption that $\Omega$ $\Omega_0$ associate with the same joint pdf \eqref{joint density 0 stage} and \eqref{joint density 1 stage}, the following identity follows immediately:
\begin{equation}
\rho_c(0,z,t\mid \Omega)\equiv \rho_c(0,z,t\mid \Omega_0)
\end{equation}
which, together with the statement (3) of lemma \ref{lemma1}, implies that
\begin{equation}\label{intensity equ}
\exp((\tilde{b}^{c,-\ast})^{\top}z)\tilde{\lambda}_0^c(t)=
\exp((b^{c,-\ast})^{\top}z)\lambda_0^c(t)
\end{equation}
\eqref{intensity equ} implies the identity, $\tilde{b}^{c,-\ast}=b^{c,-\ast}$ and $\tilde{\lambda}_0^c=\equiv\lambda_0^c$. Using statement (3) of lemma \ref{lemma1} once again, the identity in $b^{c,-\ast}$ and $\lambda_0^c$ enforces the identity $\rho_c'(0,z',z,s,t\mid\Omega)\equiv\rho_c'(0,z',z,s,t\mid\Omega)$
which furthermore enforces the identity in \eqref{jump density 0 stage} between $\Omega$ and $\Omega_0$. Consequently, the first summand in \eqref{joint density 1 stage} must be identical for $\Omega$ and $\Omega_0$. Combining it with all the identities of $a$, $b$, $\lambda_0$, $\lambda_0^c$ and $b^{c,-\ast}$, $\tilde{b}^{c,\ast}=b^{c,\ast}$ is guaranteed.

Finally, if the initial $p_Z(\cdot\mid \Omega_0)$ assign zeros mass to $Z^{\ast}(0)=0$, we can still adopt exactly the same proof as above, the only modification is replacing $j=0$ to $j'$ such that $j'$ is the smallest positive integer with $p_Z(j',\cdot,0\mid \Omega_0)>0$, such $j'$ must exist because $p_Z(\cdot,\cdot,0\mid \Omega_0)$ is a well-defined probability density function. Then, proof for lemma \ref{lemma3} is completed. 

\end{proof}


\section{Tables \& Figures}
\subsection{Tables}
\begin{minipage}{\linewidth}
\begin{center} 
\captionof{table}{Fitting Performance for $\mathcal{M}$ and $\Sigma$}\label{table: 1}
\centering
\resizebox{8cm}{4.5cm}{%
\begin{tabular}{llll|lll}
\hline
 & n=100 &  &  & n=200 &  & \tabularnewline
\hline
Var & Bias & SSE & 95\% CP & Bias & SSE & 95\% CP\tabularnewline
\hline
$\mu_{11}$ & 0.013 & 0.208 & 0.408 & 0.021 & 0.208 & 0.408\tabularnewline
$\mu_{12}$ & 0.006 & 0.218 & 0.427 & 0.006 & 0.222 & 0.435\tabularnewline
$\mu_{13}$ & -0.017 & 0.209 & 0.41 & 0.004 & 0.223 & 0.437\tabularnewline
$\mu_{14}$ & -0.002 & 0.197 & 0.386 & 0.012 & 0.201 & 0.394\tabularnewline
$\mu_{15}$ & -0.003 & 0.226 & 0.443 & 0.011 & 0.204 & 0.4\tabularnewline
$\mu_{16}$ & 0.009 & 0.203 & 0.398 & 0.014 & 0.203 & 0.398\tabularnewline
$\mu_{21}$ & -0.056 & 0.191 & 0.374 & -0.066 & 0.2 & 0.392\tabularnewline
$\mu_{22}$ & 0.039 & 0.206 & 0.404 & 0.035 & 0.191 & 0.374\tabularnewline
$\mu_{23}$ & -0.01 & 0.199 & 0.39 & -0.015 & 0.194 & 0.38\tabularnewline
$\mu_{24}$ & -0.019 & 0.199 & 0.39 & -0.003 & 0.188 & 0.368\tabularnewline
$\mu_{25}$ & 0.018 & 0.183 & 0.359 & -0.004 & 0.182 & 0.357\tabularnewline
$\mu_{26}$ & 0 & 0.215 & 0.421 & -0.026 & 0.207 & 0.406\tabularnewline
$\sigma_{11}^{2}$ & 0.118 & 0.272 & 0.533 & 0.099 & 0.234 & 0.459\tabularnewline
$\sigma_{12}^{2}$ & 0.107 & 0.244 & 0.478 & 0.085 & 0.248 & 0.486\tabularnewline
$\sigma_{13}^{2}$ & 0.097 & 0.259 & 0.508 & 0.097 & 0.214 & 0.419\tabularnewline
$\sigma_{14}^{2}$ & 0.102 & 0.26 & 0.51 & 0.075 & 0.222 & 0.435\tabularnewline
$\sigma_{15}^{2}$ & 0.096 & 0.273 & 0.535 & 0.117 & 0.238 & 0.466\tabularnewline
$\sigma_{16}^{2}$ & 0.087 & 0.249 & 0.488 & 0.109 & 0.25 & 0.49\tabularnewline
$\sigma_{21}^{2}$ & 0.053 & 0.219 & 0.429 & 0.038 & 0.219 & 0.429\tabularnewline
$\sigma_{22}^{2}$ & 0.067 & 0.211 & 0.414 & 0.025 & 0.216 & 0.423\tabularnewline
$\sigma_{23}^{2}$ & 0.077 & 0.207 & 0.406 & 0.074 & 0.219 & 0.429\tabularnewline
$\sigma_{24}^{2}$ & 0.051 & 0.222 & 0.435 & 0.041 & 0.214 & 0.419\tabularnewline
$\sigma_{25}^{2}$ & 0.072 & 0.222 & 0.435 & 0.07 & 0.205 & 0.402\tabularnewline
$\sigma_{26}^{2}$ & 0.068 & 0.224 & 0.439 & 0.039 & 0.211 & 0.414\tabularnewline
\hline\hline
\end{tabular}%
}
\end{center}
\end{minipage}\\

\vspace{1cm}
\begin{minipage}{\linewidth}
\begin{center} 
\captionof{table}{Fitting Performance for $b$ and $b^c$}\label{table: 2}
\centering
\resizebox{8cm}{3.5cm}{%
\begin{tabular}{llll|lll}
\hline\hline
 & n=100 &  &  & n=200 &  & \tabularnewline
 \hline
Var & Bias & SSE & 95\% CP & Bias & SSE & 95\% CP\tabularnewline
\hline
$b_{1}$ & 0.022 & 0.199 & 0.39 & 0.038 & 0.221 & 0.433\tabularnewline
$b_{2}$ & 0.007 & 0.199 & 0.39 & 0 & 0.211 & 0.414\tabularnewline
$b_{3}$ & 0.004 & 0.209 & 0.41 & 0.007 & 0.216 & 0.423\tabularnewline
$b_{4}$ & 0.006 & 0.206 & 0.404 & 0.024 & 0.205 & 0.402\tabularnewline
$b_{5}$ & 0.01 & 0.22 & 0.431 & 0.001 & 0.212 & 0.416\tabularnewline
$b_{6}$ & 0.006 & 0.226 & 0.443 & 0.01 & 0.218 & 0.427\tabularnewline
$b_{7}$ & 0.005 & 0.199 & 0.39 & 0.003 & 0.196 & 0.384\tabularnewline
$b_{1}^{c}$ & 0.005 & 0.227 & 0.445 & 0.022 & 0.219 & 0.429\tabularnewline
$b_{2}^{c}$ & 0.001 & 0.203 & 0.398 & -0.024 & 0.211 & 0.414\tabularnewline
$b_{3}^{c}$ & 0.007 & 0.199 & 0.39 & -0.001 & 0.206 & 0.404\tabularnewline
$b_{4}^{c}$ & 0.017 & 0.207 & 0.406 & -1.018 & 0.181 & 0.355\tabularnewline
$b_{5}^{c}$ & 0.011 & 0.2143 & 0.42 & 0.995 & 0.218 & 0.427\tabularnewline
$b_{6}^{c}$ & 0.003 & 0.229 & 0.449 & 0.623 & 0.227 & 0.445\tabularnewline
$b_{7}^{c}$ & 0.022 & 0.232 & 0.455 & 0.991 & 0.206 & 0.404\tabularnewline
\hline\hline
\end{tabular}%
}
\end{center}
\end{minipage}\\

\vspace{1cm}
\begin{minipage}{\linewidth}
\begin{center} 
\captionof{table}{Estimated $b$ and $b^c$ for renrendai data}\label{table: 3}
\centering
\resizebox{7cm}{4.5cm}{%
\begin{tabular}{lll}
\hline\hline
Var & $b$ & $b^{c}$\tabularnewline
\hline
$Z_{1}$ Term & 0 & 0\tabularnewline
$Z_{2}$ Interest Rate & 0 & 0\tabularnewline
$Z_{3}$ Principal & 0 & 0\tabularnewline
$Z_{4}$ Age & 0 & 0\tabularnewline
$Z_{5}$ Credit Score & 0 & 0\tabularnewline
$Z_{6}$ Education & 0 & 0\tabularnewline
$Z_{7}$ Income & 0 & 0\tabularnewline
$Z_{8}$ Married & 0 & 0\tabularnewline
$Z_{9}$ Divorce & 0 & -0.157\tabularnewline
$Z_{10}$Unpaid Car Loan  & 0 & 0.012\tabularnewline
$Z_{11}$Car Owned & 0 & 0\tabularnewline
$Z_{12}$Unpaid Mortgage & 0 & 0\tabularnewline
$Z_{13}$House Owned & 0 & 0\tabularnewline
$Z_{14}$Clerk & 0.088 & 0.006\tabularnewline
$Z_{15}$Self Employed & 0 & 0.091\tabularnewline
$Z_{16}$Business Owner/Manager & 0 & 0.298\tabularnewline
$Z_{17}$Company Scale & 0 & 0.02\tabularnewline
$Z_{18}$Local GDP & 0.617 & -0.066\tabularnewline
$Z_{19}$Local Housing Price & 0.667 & 0\tabularnewline
$Z_{20}$Irregular Payments & -0.037 & 0\tabularnewline
\hline\hline
\end{tabular}%
}
\end{center}
\end{minipage}\\
\subsection{Figures}
\begin{minipage}\linewidth
\begin{center}
\includegraphics[width=15cm,height=16cm]{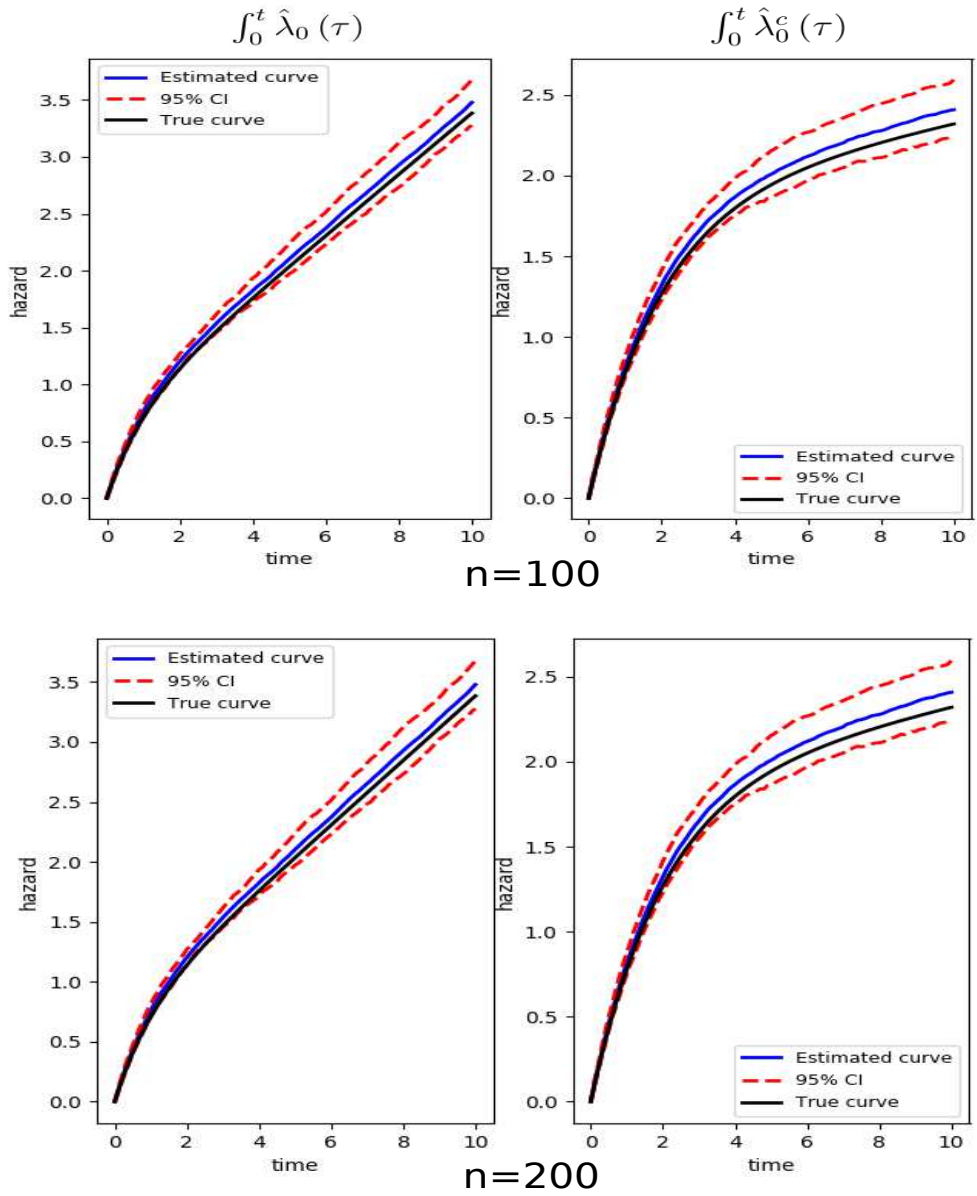}
\captionof{figure}{Estimated $\int_{0}^{t}\hat{\lambda}_0(\tau)d\tau$, $\int_{0}^{t}\hat{\lambda}_0^c(\tau)d\tau$ v.s. True $\int_{0}^{t}\lambda_0(\tau)d\tau$, $\int_{0}^{t}\lambda_0^c(\tau)d\tau$}\label{fig: fitting}
%
\end{center}
\end{minipage}\\

\vspace{1cm}
\begin{minipage}\linewidth
\begin{center}
\includegraphics[width=15cm,height=10cm]{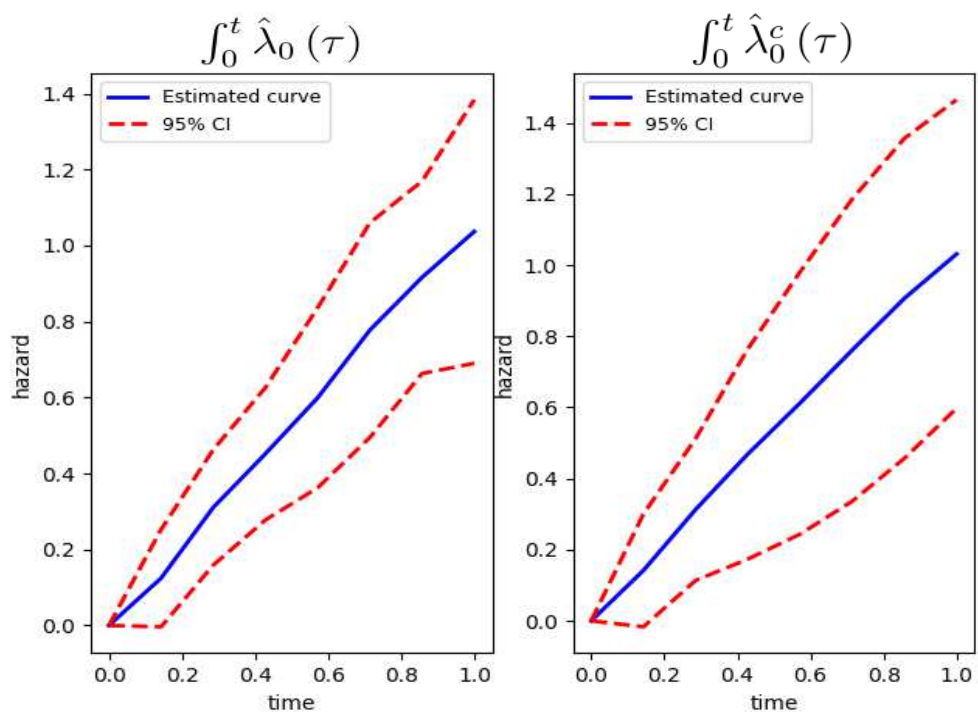}
\captionof{figure}{Estimated $\int_{0}^{t}\hat{\lambda}_0(\tau)d\tau$, $\int_{0}^{t}\hat{\lambda}_0^c(\tau)d\tau$ for renrendai data}\label{fig: real}
%
\end{center}
\end{minipage}\\

\begin{thebibliography}{36}
\providecommand{\natexlab}[1]{#1}
\providecommand{\url}[1]{\texttt{#1}}
\expandafter\ifx\csname urlstyle\endcsname\relax
  \providecommand{\doi}[1]{doi: #1}\else
  \providecommand{\doi}{doi: \begingroup \urlstyle{rm}\Url}\fi

\bibitem[Amemiya(1985)]{amemiya1985advanced}
Amemiya~T.
\newblock \emph{Advanced econometrics}.
\newblock Harvard university press, 1985.

\bibitem[Andersen(1992)]{andersen1992repeated}
Andersen~P.K.
\newblock Repeated assessment of risk factors in survival analysis.
\newblock \emph{Statistical Methods in Medical Research}, 1\penalty0
  (3):\penalty0 297--315, 1992.

\bibitem[Andersen and Gill(1982)]{andersen1982cox}
Andersen~P.K. and Gill~R.D.
\newblock Cox's regression model for counting processes: a large sample study.
\newblock \emph{The annals of statistics}, pages 1100--1120, 1982.

\bibitem[Barrett et~al.(2015)]{barrett2015joint}
Barrett~J., Diggle~P., Henderson~R. and Taylor-Robinson~D.
\newblock Joint modelling of repeated measurements and time-to-event outcomes:
  flexible model specification and exact likelihood inference.
\newblock \emph{Journal of the Royal Statistical Society: Series B (Statistical
  Methodology)}, 77\penalty0 (1):\penalty0 131--148, 2015.

\bibitem[Cauwenberghs(1993)]{cauwenberghs1993fast}
Cauwenberghs~G.
\newblock A fast stochastic error-descent algorithm for supervised learning and
  optimization.
\newblock In \emph{Advances in neural information processing systems}, pages
  244--251, 1993.

\bibitem[Chen et~al.(2014)]{chen2014joint}
Chen~Q, May~R.C., Ibrahim~J.G., Chu~H. and Cole~S.R.
\newblock Joint modeling of longitudinal and survival data with missing and
  left-censored time-varying covariates.
\newblock \emph{Statistics in medicine}, 33\penalty0 (26):\penalty0 4560--4576,
  2014.

\bibitem[Cox(1972)]{cox1972regression}
Cox~D.R.
\newblock Regression models and life-tables (with discussion).
\newblock In \emph{J. R. Statist. Soc.}, volume~B, pages 187--220. 1972.

\bibitem[Guo(2012)]{guo2012parallel}
Guo~G.
\newblock Parallel statistical computing for statistical inference.
\newblock \emph{Journal of Statistical Theory and Practice}, 6\penalty0
  (3):\penalty0 536--565, 2012.

\bibitem[Guo and Carlin(2004)]{guo2004separate}
Guo~X. and Carlin~B.P.
\newblock Separate and joint modeling of longitudinal and event time data using
  standard computer packages.
\newblock \emph{The American Statistician}, 58\penalty0 (1):\penalty0 16--24,
  2004.

\bibitem[Hogan and Laird(1997)]{hogan1997model}
Hogan~J.W. and Laird~N.M.
\newblock Model-based approaches to analysing incomplete longitudinal and
  failure time data.
\newblock \emph{Statistics in medicine}, 16\penalty0 (3):\penalty0 259--272,
  1997.

\bibitem[Karatzas and Shreve(2012)]{karatzas2012brownian}
Karatzas~I. and Shreve~S.
\newblock \emph{Brownian motion and stochastic calculus}, volume 113.
\newblock Springer Science \& Business Media, 2012.

\bibitem[Karr(2017)]{karr2017point}
Karr~A.
\newblock \emph{Point processes and their statistical inference}.
\newblock Routledge, 2017.

\bibitem[Kim et~al.(2013)]{kim2013joint}
Kim~S., Zeng~D., Li~Y. and Spiegelman~D.
\newblock Joint modeling of longitudinal and cure-survival data.
\newblock \emph{Journal of statistical theory and practice}, 7\penalty0
  (2):\penalty0 324--344, 2013.

\bibitem[Koopman et~al.(2008)]{koopman2008multi}
Koopman~S.J., Lucas~A. and Monteiro~A.
\newblock The multi-state latent factor intensity model for credit rating
  transitions.
\newblock \emph{Journal of Econometrics}, 142\penalty0 (1):\penalty0 399--424,
  2008.

\bibitem[Laird(1988)]{laird1988missing}
Laird~N.M.
\newblock Missing data in longitudinal studies.
\newblock \emph{Statistics in medicine}, 7\penalty0 (1-2):\penalty0 305--315,
  1988.

\bibitem[Li et~al.(2017)]{li2017impact}
Li~Z., Shao~A.W. and Sherris~M.
\newblock The impact of systematic trend and uncertainty on mortality and
  disability in a multistate latent factor model for transition rates.
\newblock \emph{North American Actuarial Journal}, 21\penalty0 (4):\penalty0
  594--610, 2017.

\bibitem[Lin et~al.(2000)]{lin2000semiparametric}
Lin~D., Wei~L., Yang~I. and Ying~Z.
\newblock Semiparametric regression for the mean and rate functions of
  recurrent events.
\newblock \emph{Journal of the Royal Statistical Society: Series B (Statistical
  Methodology)}, 62\penalty0 (4):\penalty0 711--730, 2000.

\bibitem[Liu et~al.(2015)]{liu2015asynchronous}
Liu~J, Wright~S.J., R{\'e}~C.,  Bittorf~V. and 
Sridhar~S.
\newblock An asynchronous parallel stochastic coordinate descent algorithm.
\newblock \emph{The Journal of Machine Learning Research}, 16\penalty0
  (1):\penalty0 285--322, 2015.

\bibitem[Osmera et~al.(2003)]{osmera2003parallel}
Osmera~P., Lacko~B. and Petr~M.
\newblock Parallel evolutionary algorithms.
\newblock In \emph{Computational Intelligence in Robotics and Automation, 2003.
  Proceedings. 2003 IEEE International Symposium on}, volume~3, pages
  1348--1353. IEEE, 2003.

\bibitem[Perko(2013)]{perko2013differential}
Perko~L.
\newblock \emph{Differential equations and dynamical systems}, volume~7.
\newblock Springer Science \& Business Media, 2013.

\bibitem[Riphahn et~al.(2003)]{riphahn2003incentive}
Riphahn~R.T., Wambach~A. and Million~A.
\newblock Incentive effects in the demand for health care: a bivariate panel
  count data estimation.
\newblock \emph{Journal of applied econometrics}, 18\penalty0 (4):\penalty0
  387--405, 2003.

\bibitem[Rizopoulos(2010)]{rizopoulos2010jm}
Rizopoulos~D.
\newblock Jm: An r package for the joint modelling of longitudinal and
  time-to-event data.
\newblock \emph{Journal of Statistical Software (Online)}, 35\penalty0
  (9):\penalty0 1--33, 2010.

\bibitem[Rizopoulos(2011)]{rizopoulos2011dynamic}
Rizopoulos~D.
\newblock Dynamic predictions and prospective accuracy in joint models for
  longitudinal and time-to-event data.
\newblock \emph{Biometrics}, 67\penalty0 (3):\penalty0 819--829, 2011.

\bibitem[Sattar and Sinha(2017)]{sattar2017joint}
Sattar~A. and Sinha~S.K.
\newblock Joint modeling of longitudinal and survival data with a covariate
  subject to a limit of detection.
\newblock \emph{Statistical methods in medical research}, page
  0962280217729573, 2017.

\bibitem[Sudholt(2015)]{sudholt2015parallel}
Sudholt~D.
\newblock Parallel evolutionary algorithms.
\newblock In \emph{Springer Handbook of Computational Intelligence}, pages
  929--959. Springer, 2015.

\bibitem[Sun(2014)]{sun2014panel}
Sun~J.
\newblock Panel count data.
\newblock \emph{Wiley StatsRef: Statistics Reference Online}, 2014.

\bibitem[Taylor et~al.(1994)]{taylor1994stochastic}
Taylor~J.M., Cumberland~W.G. and Sy~J.P.
\newblock A stochastic model for analysis of longitudinal aids data.
\newblock \emph{Journal of the American Statistical Association}, 89\penalty0
  (427):\penalty0 727--736, 1994.

\bibitem[Tibshirani(1996)]{tibshirani1996regression}
Tibshirani~R.
\newblock Regression shrinkage and selection via the lasso.
\newblock \emph{Journal of the Royal Statistical Society. Series B
  (Methodological)}, pages 267--288, 1996.

\bibitem[Tomassini(1999)]{tomassini1999parallel}
Tomassini~M.
\newblock Parallel and distributed evolutionary algorithms: A review.
\newblock 1999.

\bibitem[Tsiatis et~al.(1995)]{tsiatis1995modeling}
Tsiatis~A.A., Degruttola~V. and Wulfsohn~M.S.
\newblock Modeling the relationship of survival to longitudinal data measured
  with error. applications to survival and cd4 counts in patients with aids.
\newblock \emph{Journal of the American Statistical Association}, 90\penalty0
  (429):\penalty0 27--37, 1995.

\bibitem[Wang and Taylor(2001)]{wang2001jointly}
Wang~Y. and Taylor~J.M.
\newblock Jointly modeling longitudinal and event time data with application to
  acquired immunodeficiency syndrome.
\newblock \emph{Journal of the American Statistical Association}, 96\penalty0
  (455):\penalty0 895--905, 2001.

\bibitem[Wu and Yu(2014)]{wu2014joint}
Wu~L. and Yu~T.
\newblock Joint modeling of longitudinal and survival data.
\newblock \emph{Wiley StatsRef: Statistics Reference Online}, pages 1--9, 2014.

\bibitem[Wu et~al.(2007)]{wu2007joint}
Wu~L., Hu~X.J. and Wu~H.
\newblock Joint inference for nonlinear mixed-effects models and time to event
  at the presence of missing data.
\newblock \emph{Biostatistics}, 9\penalty0 (2):\penalty0 308--320, 2007.

\bibitem[Zeng and Lin(2007)]{zeng2007maximum}
Zeng~D. and Lin~D.Y.
\newblock Maximum likelihood estimation in semiparametric regression models
  with censored data.
\newblock \emph{Journal of the Royal Statistical Society: Series B (Statistical
  Methodology)}, 69\penalty0 (4):\penalty0 507--564, 2007.

\bibitem[Zheng et~al.(2018)]{zheng2018understanding}
Zheng~Y., Zhao~X. and Zhang~X.
\newblock Understanding dynamic status change of hospital stay and cost
  accumulation via combining continuous and finitely jumped processes.
\newblock \emph{Computational and Mathematical Methods in Medicine}, 2018,
  2018.

\bibitem[Zou(2006)]{zou2006adaptive}
Zou~H.
\newblock The adaptive lasso and its oracle properties.
\newblock \emph{Journal of the American statistical association}, 101\penalty0
  (476):\penalty0 1418--1429, 2006.

\end{thebibliography}

\end{appendix}

\end{document}